\documentclass[a4paper,fleqn,usenatbib,useAMS]{mnras}

\usepackage{graphicx}	
\usepackage{amsmath}	
\usepackage{amssymb}	
\usepackage{multicol}        
\usepackage{bm}		
\usepackage{pdflscape}	

\usepackage[T1]{fontenc}
\usepackage{ae,aecompl}

\usepackage{newtxtext,newtxmath}

\title[Massive PSBs at $z>1$ are compact
  proto-spheroids]{Massive post-starburst galaxies at $z>1$ are
  compact proto-spheroids}

\author[O. Almaini et al.]{Omar Almaini$^{1}$\thanks{E-mail:
    omar.almaini@nottingham.ac.uk}, Vivienne Wild$^{2}$, David T.
  Maltby$^{1}$, William G. Hartley$^{3}$, \newauthor 
Chris Simpson$^{4}$, 
  Nina A. Hatch$^{1}$,  
Ross J. McLure$^{5}$, James S. Dunlop$^{5}$,
  Kate Rowlands$^{2}$ \\ $^{1}$School of Physics and Astronomy,
  University of Nottingham, University Park, Nottingham NG7 2RD,
  U.K. \\ $^2$School of Physics and Astronomy, University of St
  Andrews, North Haugh, St Andrews, KY16 9SS, U.K. \\ $^3$ETH
  Z\"urich, Institut f\"ur Astronomie, HIT J 11.3,
  Wolfgang-Pauli-Str. 27, 8093 Z\"urich, Switzerland 
\\ $^4$Gemini Observatory, Northern Operations Center, 670 N. A'ohuku Place, Hilo, HI96720, USA
\\ $^5$Institute
  for Astronomy, University of Edinburgh, Royal Observatory, Blackford
  Hill, Edinburgh, EH9 3HJ, U.K. \\ }

\date{Accepted 2017 July 28. Received 2017 July 24; in original form 2016 February 3.}

\pubyear{2016}

\begin{document}
\label{firstpage}
\pagerange{\pageref{firstpage}--\pageref{lastpage}}
\maketitle


\begin{abstract}

We investigate the relationship between the quenching of star
formation and the structural transformation of massive galaxies, using
a large sample of photometrically-selected post-starburst galaxies in
the UKIDSS UDS field. We find that post-starburst galaxies at
high-redshift ($z>1$) show high S\'{e}rsic indices, significantly
higher than those of active star-forming galaxies, but with a
distribution that is indistinguishable from the old quiescent
population.  We conclude that the morphological transformation occurs
before (or during) the quenching of star formation.  Recently quenched
galaxies are also the most compact; we find evidence that massive
post-starburst galaxies (M$_{\ast}> 10^{10.5} ~$M$_{\sun}$) at high
redshift ($z>1$) are  on average smaller than comparable quiescent
galaxies at the same epoch.  Our findings are consistent with a
scenario in which massive passive galaxies are formed from three
distinct phases: (1) gas-rich dissipative collapse to very high
densities, forming the proto-spheroid; (2) rapid quenching of star
formation, to create the ``red nugget'' with post-starburst features;
(3) a gradual growth in size as the population ages, perhaps as a
result of minor mergers.

\end{abstract}




\section{Introduction}
\label{sec:intro}

Galaxies in the local Universe display a striking bimodality in their
morphological and spectral characteristics; massive galaxies
(M$_{\ast}> 10^{10.5} ~$M$_{\sun}$) are typically spheroidal with old
stellar populations, while lower mass galaxies are typically
disc-dominated with blue, younger stellar populations (e.g. Strateva
et al. 2001; Hogg et al. 2002).  Deep surveys have revealed that the
most massive galaxies were formed at high redshift ($z>1$;
e.g. Fontana et al. 2004; Kodama et al. 2004; Cirasuolo et al. 2010),
but we still do not understand why their star formation was abruptly
terminated. Feedback from AGN (e.g. Silk \& Rees 1998; Hopkins et
al. 2005) or starburst-driven superwinds (e.g. Diamond-Stanic et
al. 2012) are leading contenders for rapidly quenching distant
galaxies, while jet-mode AGN feedback may be required to maintain the
``red and dead'' phase (e.g. Best et al. 2006).

In addition to the quenching of star formation, massive galaxies must
also undergo a dramatic structural transformation, to
produce the spheroid-dominated population we see today.  The
transition appears to occur at $z>1$ for most galaxies with M$_{\ast}>
10^{10.5} ~$M$_{\sun}$ (Mortlock et al. 2013; Bruce et al. 2014), but
it is unclear if quenching and structural transformation occurred
during the same event.  Over the last 10 years it has also emerged
that quiescent galaxies were significantly more compact in the early
Universe compared to the present day (e.g. Daddi et al. 2005; Trujillo
et al. 2006; van Dokkum et al. 2008; Belli et al. 2014).  
Plausible explanations for the dramatic size growth include minor mergers
(e.g. Bezanson et al. 2009; Naab, Johansson \& Ostriker
2009) or  expansion due to mass loss (Fan et al. 2008). For the population
of quiescent galaxies as a whole, however, there may also  be an
element of progenitor bias; galaxies quenched at lower redshift tend
to be larger than their counterparts at early times, which may drive
much of the observed size evolution (e.g. Poggianti et al. 2013;
Carollo et al. 2013).

From a theoretical perspective, the formation of ultra-compact massive
spheroids requires the concentration of vast reservoirs of cool gas
via dissipation, which can radiate and collapse to very high
densities. A variety of models have arisen to explain spheroid
formation in detail, with some invoking gas-rich mergers (e.g. Hopkins
et al. 2009; Wellons et al. 2015) while others use the inflow of gas
through cold streams, feeding an extended disc that eventually becomes
unstable and contracts (e.g. Dekel et al. 2009; Zolotov et
al. 2015). Outflows driven by AGN or star formation may then terminate
the star formation by expelling the remainder of the gas (e.g. Hopkins
et al. 2005).

In this work we explore the relationship between the quenching of
distant galaxies and their structural transformation.  We focus in
particular on the rare class of ``post-starburst'' (PSB) galaxies,
which are observed a few hundred Myr after a major episode of star
formation was rapidly quenched.  In the local Universe PSBs are
identified from characteristic strong Balmer absorption lines
(Dressler \& Gunn 1983; Wild et al. 2009), due to the strong
contribution from A stars, but until recently very few were
spectroscopically identified at $z>1$ (e.g.  Vergani et al. 2010).
Two photometric methods have therefore been developed to identify this
population. Whitaker et al.  (2012) used medium-band photometry from
the NEWFIRM Medium-Band Survey to identify ``young red-sequence''
galaxies using rest-frame UVJ colour-colour diagrams.  In Wild et al.
(2014) an alternative approach was used, based on a Principal Component
Analysis (PCA) applied to the deep multi-wavelength photometry in the
UDS field.  Three spectral shape parameters (``supercolours'') were
found to provide a compact representation for a wide range of
photometric SEDs.  In addition to cleanly separating quiescent and
star-forming galaxies, the PCA method identifies ``post-starburst''
galaxy candidates in a distinct region of supercolour space,
corresponding to galaxies in which a significant amount of mass was
formed within the last Gyr but then rapidly quenched.  The method was
recently verified with deep 8m spectroscopy from VLT, which
established that between 60\% and 80\% of photometric candidates are
spectroscopically confirmed post-starburst galaxies, depending on the
specific criteria adopted (Maltby et al. 2016)\footnote{From a sample
  of 24 PSB candidates, 19 galaxies ($\sim$80\%) showed strong Balmer
  absorption lines ($W_{\rm H\delta}>5$\AA), dropping to 14
  confirmations ($\sim$60\%) if stricter criteria are used to exclude
  galaxies with significant [O\,\textsc{ii}] emission.  The fraction
  of spectroscopic PSBs among the passive and star-forming PCA classes
  was estimated to be $<$10\% and $<$1\% respectively (Maltby; private
  communication).}.  In terms of completeness, the photometric method
was found to identify approximately 60\% of galaxies that would be
spectroscopically classified as PSBs.  Overall, these figures confirm
that photometric PCA techniques can be used to identify large and
relatively clean samples of recently quenched galaxies.  So far the
spectroscopic confirmation is restricted to $z<1.4$, given the
requirement to detect H\,$\delta$ with optical spectroscopy.  Future
near-infrared spectroscopy (e.g. with the {\it James Webb Space
  Telescope}) will allow a detailed investigation of the population at
higher redshift.

The identification  of large PSB samples has allowed the first
study of  the PSB galaxy mass function and its evolution to
$z=2$ (Wild et al. 2016). Strong evolution was observed, with the
implication that a large fraction of massive galaxies are rapidly
quenched and pass through a PSB phase.  In this paper we use the
unique PCA sample described in Wild et al. (2016) to explore the
structural properties of post-starburst galaxies.  As newly quenched
systems, our primary aim is to investigate if this population is
structurally similar to star-forming galaxies at the same epoch, or if
they already show evidence for the compact spheroid-dominated
morphology of well-established quiescent galaxies.

We assume a cosmology with $\Omega_M=0.3$, $\Omega_\Lambda=0.7$ and
$h=0.7$.  All magnitudes are given in the AB system.

\section{Data and sample selection}

\subsection{The UDS K-band galaxy sample}

Our study is based on deep $K$-band imaging from the UKIRT Infrared Deep
Sky Survey (UKIDSS; Lawrence et al. 2007) Ultra-Deep Survey (UDS;
Almaini et al., in preparation). The UDS is the deepest of the UKIDSS surveys,
covering $0.77$ square degrees in the $J$, $H$ and $K$ bands.  We use
the 8th UDS data release (Hartley et al. 2013), reaching depths of
$J=24.9$, $H=24.2$ and $K=24.6$ (AB, 5$\sigma$, 2-arcsec
apertures). The final UDS data release (June 2016) achieved
estimated depths $J=25.4, H=24.8, K=25.3$, and will be used to extend
our PSB studies in future work.

To complement the near-infrared imaging from UKIDSS, the UDS has deep
optical coverage from {\it Subaru} Suprime-CAM, to depths of $B= 27.6,
~ V= 27.2, ~ R = 27.0, ~ i^{\prime} = 27.0$ and $ z^{\prime} = 26.0$
(AB, 5$\sigma$, $2$ arcsec), as described in Furusawa et
al. (2008). Additional $u^{\prime}$-band imaging is provided by the
Canada-France-Hawaii Telescope (CFHT) MegaCam instrument, reaching
$ u^{\prime} = 26.75$ (AB, 5$\sigma$, $2$ arcsec). Deep imaging
at longer near-infrared wavelengths is provided by the {\it Spitzer}
UDS Legacy Program (SpUDS; PI: Dunlop), achieving depths of 24.2 and
24.0 (AB) using the IRAC camera at 3.6$\mu$m and 4.5$\mu$m
respectively.  The resulting area with full multiwavelength coverage,
following the masking of bright stars and artefacts, is 0.62 square
degrees. Further details on the construction of the multiwavelength
DR8 catalogue can be found in Hartley et al. (2013).

We determined photometric redshifts using the techniques outlined in
Simpson et al. (2013). The 11-band photometric data were fit using a
grid of galaxy templates, assembled using simple stellar populations
from Bruzual \& Charlot (2003; hereafter BC03), with a logarithmic
spacing of ages between 30 Myr and 10 Gyr, and the addition of younger
templates with dust-reddened spectral energy distributions (SEDs). The
additional templates consist of a mildly reddened ($A_V=0.25$ mag)
version of the two youngest templates, plus a version of the 30 Myr
template with heavier reddening ($A_V=1.0$ mag).  The resulting
photometric redshifts show a normalized median absolute deviation
$\sigma_{{\rm NMAD}} = 0.027$. The stellar masses used in this work
differ slightly from those presented in Simpson et al. (2013), and
instead are based on supercolour templates (see Section
\ref{sec:masses}). As noted in Section \ref{sec:uncertainties}, 
we investigated a range of alternative stellar masses, including those
from Simpson et al. (2013), and found no significant impact on our conclusions.

\subsection{Classification using PCA supercolours}

We classify galaxies using a Principal Component Analysis (PCA) method
applied to the broad-band photometric data, using the techniques
outlined in Wild et al. (2014; hereafter W14). The aim of the PCA
analysis is to describe the variation in galaxy SEDs using the linear
combination of only a small number of components. The components are
derived using a library of spectral synthesis models from BC03, using
a wide range of stochastic star formation histories, metallicities,
and dust-reddening. We found that only three principal components
(effectively low-resolution ``eigenspectra'') are required, in linear
combination, to account for $>$99.9\% of the variance in photometric
SEDs. The amplitude of each component defines a ``supercolour'',
analogous to a traditional colour but defined using all available
photometric bands.  Supercolours allow the comparison of SEDs without
the need for model fitting, and galaxies with extreme properties are
free to have colours that differ substantially from any of the input
model components.

Using the supercolour technique, we separated the UDS galaxy
population into three categories; passive galaxies (with low specific
star-formation rates), star-forming galaxies, and post-starburst
galaxies. Post-starburst galaxies are identified as a well-defined
stream of galaxies in supercolour space, consistent with quiescent
stellar populations in which a large fraction ($>$10\%) of the stellar
mass was formed within the last $\sim$1~Gyr, with the star formation
then rapidly quenched (W14; see also Wild et al. 2016). The precise
boundary between the passive and post-starburst population is set by
the ability to observe strong Balmer absorption lines in optical
spectroscopy (W14; Maltby et al. 2016). As outlined in Wild et
al. (2016), evolutionary tracks suggest that not all photometric PSBs
may have necessarily undergone a short-lived `burst' of star
formation.  The population may also include galaxies that have
undergone more extended ($\la$ 3~Gyr) periods of star formation, but
the key characteristic is the rapid quenching of star formation within
the last $\sim$1~Gyr.

In W14, four categories of star-forming galaxies were identified (SF1,
SF2, SF3 and dusty), which we combine for the purposes of this work.
As outlined in W14 (see also Section \ref{sec:uvj}) the classification by
supercolours is in very good agreement with the separation of galaxies
using the more traditional rest-frame UVJ technique for separating
star-forming and passive galaxies (Labb\'e et al. 2005; Wuyts et
al. 2007). There is also good agreement with the UVJ method of
Whitaker et al. (2012), who identified recently-quenched candidates at
the blue end of the passive UVJ sequence.

\begin{figure}
	\includegraphics[width=0.9\linewidth]{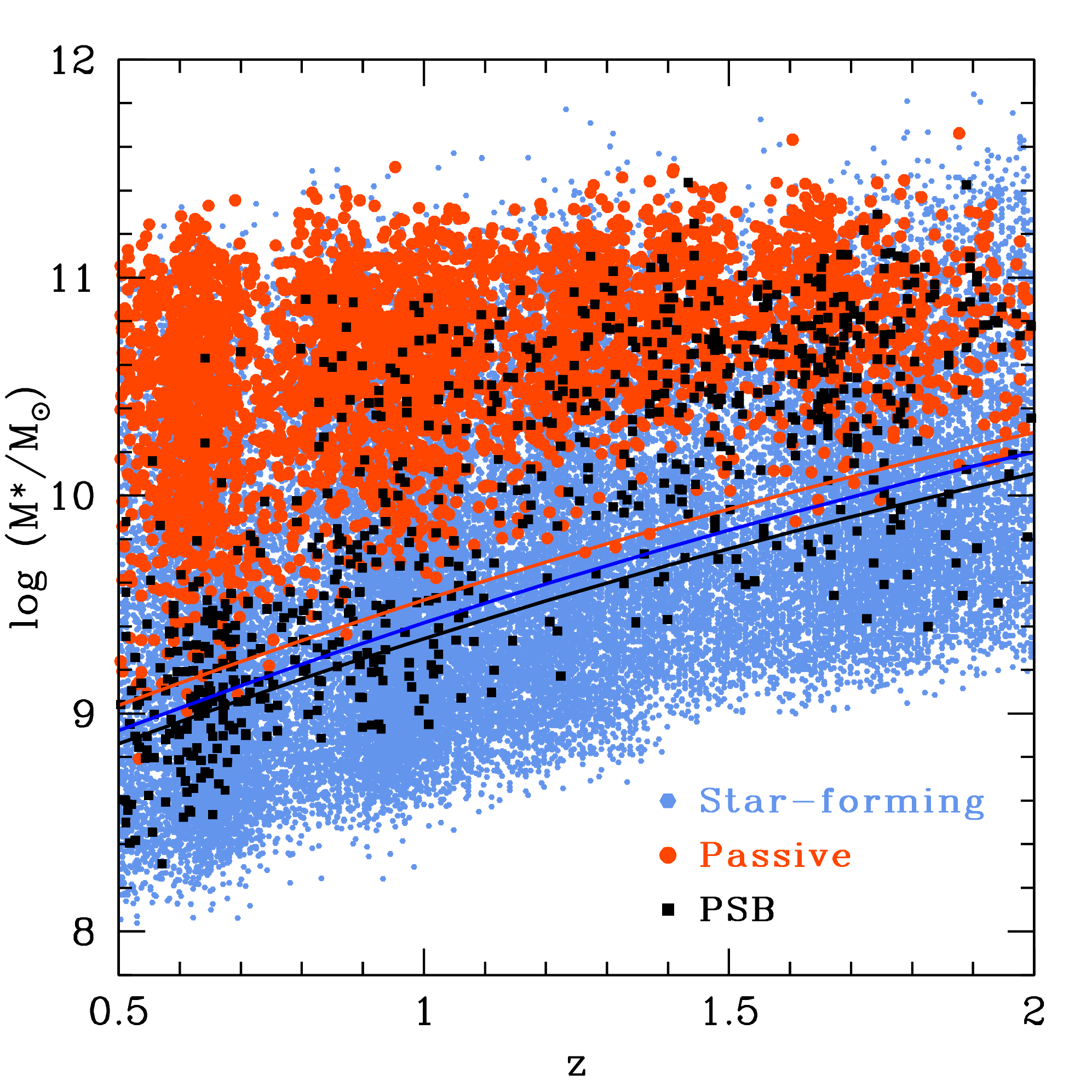}
    \caption{The distribution of stellar mass as a function of
      photometric redshift for the three primary galaxy
      populations. The curves show the corresponding 95\% completeness
      limits, determined using the method of Pozzetti et
      al. (2010). Details of the galaxy classification and stellar
      mass determination can be found in Wild et al. (2016). In this
      paper we focus on the structural properties of galaxies in the
      redshift range $1<z<2$.}
    \label{fig:zmass}
\end{figure}

\begin{figure}
	\includegraphics[width=0.9\linewidth]{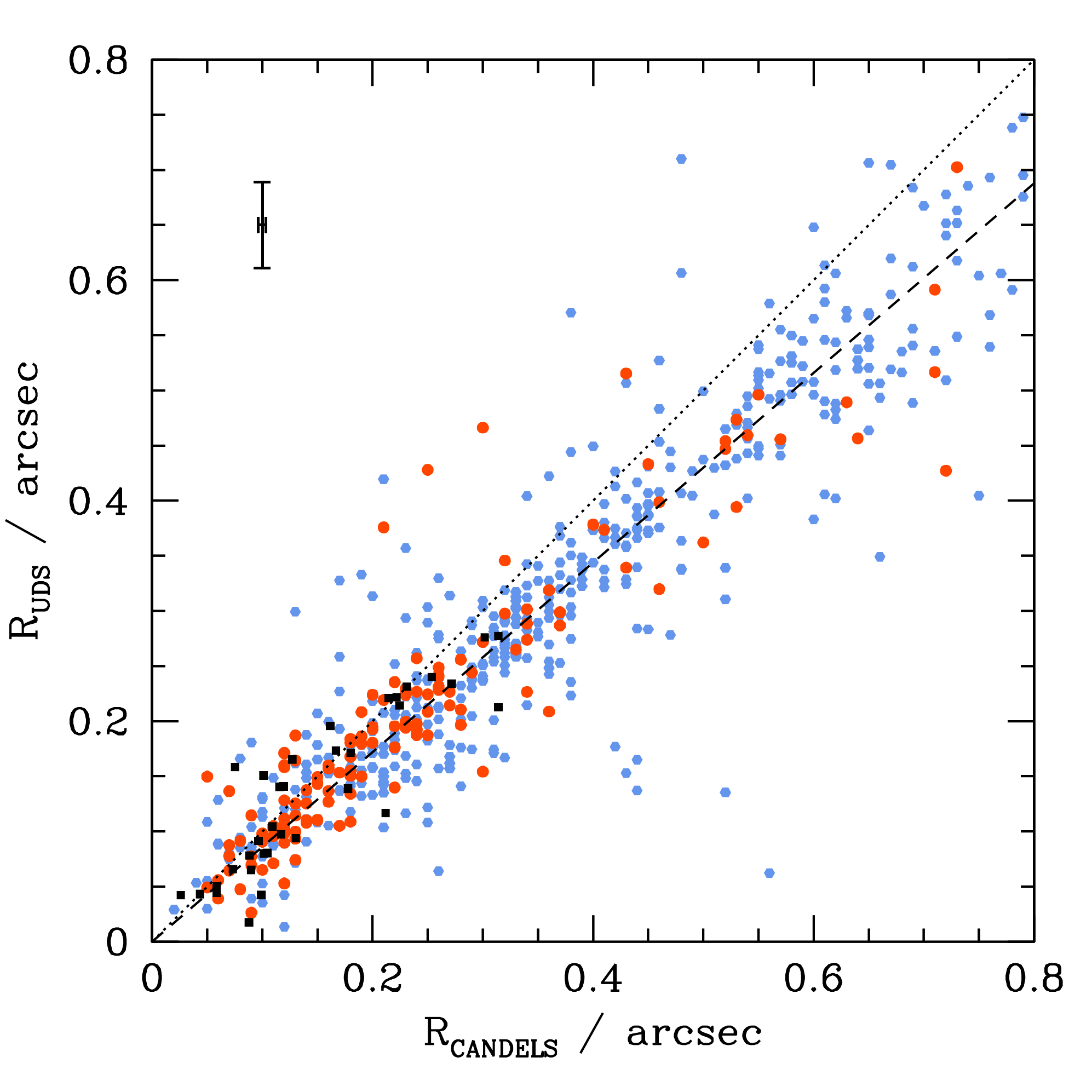}
        \includegraphics[width=0.9\linewidth]{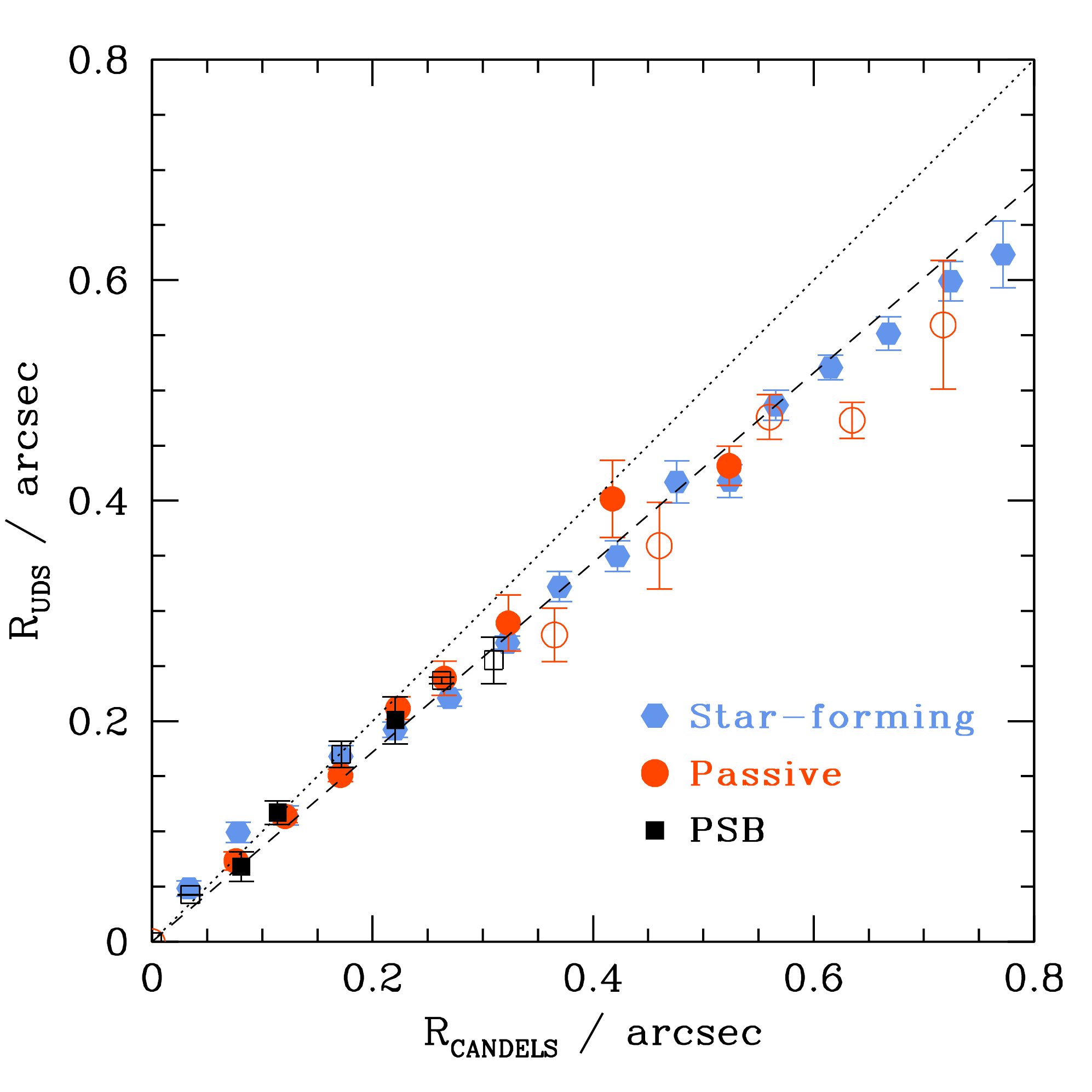}
    \caption{A comparison of size measurements for galaxies measured
      with ground-based $K$-band imaging and HST $H$-band imaging,
      using the subsample of $\sim$9\% of galaxies within the UDS
      CANDELS mosaic. The upper panel compares half-light radii for
      individual galaxies. We find that ground-based imaging can
      provide robust measurements of galaxy sizes, although the
      relationship is not precisely 1:1. A linear fit to all points
      suggests that ground-based sizes are $\sim$13\% smaller on
      average (dashed line).  A characteristic uncertainty on
      individual measurements is shown, representing the median error
      on the PSB sizes.  The lower (binned) distribution compares mean
      sizes for the three key populations, with unfilled symbols
      representing mean values determined from fewer than 5 galaxies.
      We find no evidence for a significant systematic bias in
      ground-based size determinations with spectral type. Galaxies
      shown have photometric redshifts in the range $1<z<2$ and
      stellar masses with log (M$_{\ast}$/M$_{\sun}$)$>10$, to match
      the samples used in this work.}
    \label{fig:size-test}
\end{figure}

\subsection{Stellar masses}
\label{sec:masses}

We determined stellar masses for each galaxy using a Bayesian
analysis, to account for the degeneracy between physical parameters.
Further details may be found in Wild et al. (2016).  A library of 10's
of thousands of population synthesis models was created using BC03 and
fit to the supercolours to obtain a probability density function for
each physical property. A wide range of star formation histories, dust
properties, and metallicities were explored, including exponentially
declining star formation rates with superimposed stochastic
starbursts. Stellar masses were calculated assuming a Chabrier initial
mass function (IMF), defined as the stellar mass at the time of
observation, i.e. allowing for the fraction of mass in stars returned
to the interstellar medium due to mass loss and supernovae.  The
resulting stellar mass uncertainties from the Bayesian fits are
typically $\pm ~0.1$ dex for all populations,
assuming BC03 stellar
population synthesis models and allowing for uncertainties in the
photometric redshifts. 
The potential impact of stellar mass errors
(random and systematic) is discussed further in Section
\ref{sec:uncertainties}.

Figure \ref{fig:zmass} shows the resulting distribution of
stellar mass as a function of redshift, separated into the three
primary populations. 
Mass completeness limits (95\%) were determined
as a function of redshift using the method of Pozzetti et
al. (2010). We note that the 95\% completeness limit for star-forming
galaxies appears to be surprisingly high in this diagram (formally
slightly higher than the PSB population). This is caused by the 
wide range in mass-to-light ratios within the star-forming population.

As outlined in Wild et al. (2016), the PSB mass function evolves
strongly with redshift, so that the comoving space density of massive
PSBs (M$_{\ast}> 10^{10} ~$M$_{\sun}$) is several times higher at
$z\sim 2$ than at $z\sim 0.5$. This trend is apparent in Figure
\ref{fig:zmass}, which shows a sharp decline in the number of massive
PSBs at $z<1$. The majority of PSBs at $z<1$ are close to the 95\%
completeness limit (see Figure \ref{fig:zmass}) and typically very
faint; the median $K$-band magnitude for PSBs at $z<1$ is $K=23.0$,
compared to $K=21.8$ at $z>1$.  In this work we therefore concentrate
on the structural properties of PSBs at $z>1$, and defer an
examination of the low-redshift ground-based sample to future work
using deeper $K$-band imaging. In the redshift range $1<z<2$, our
initial sample consists of 24,880 star-forming galaxies, 2043 passive
galaxies and 502 PSBs, of which 9183, 2001 and 385, respectively, have
stellar masses M$_{\ast}> 10^{10} ~$M$_{\sun}$.

\begin{figure}
	\includegraphics[width=0.9\linewidth]{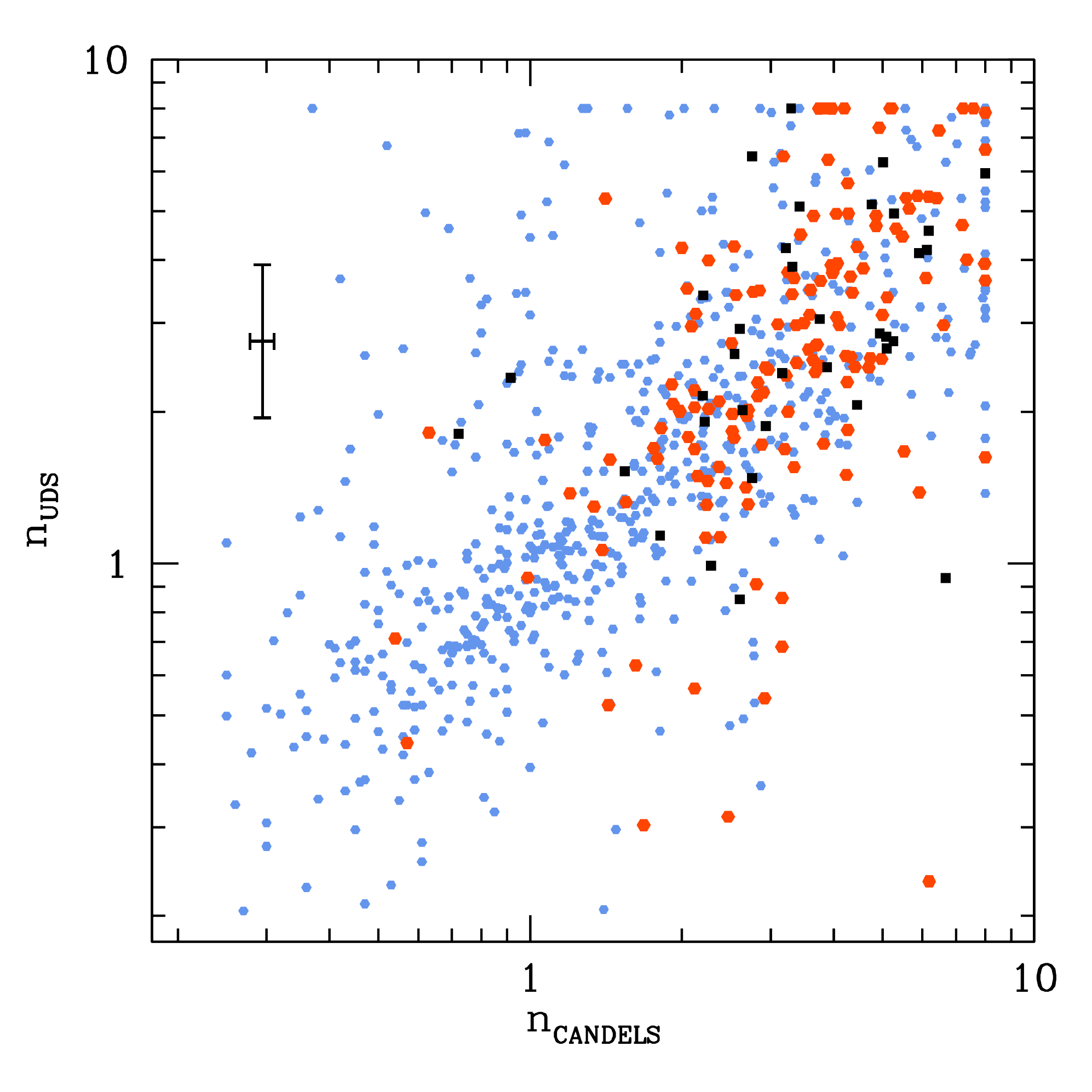}
	\includegraphics[width=0.9\linewidth]{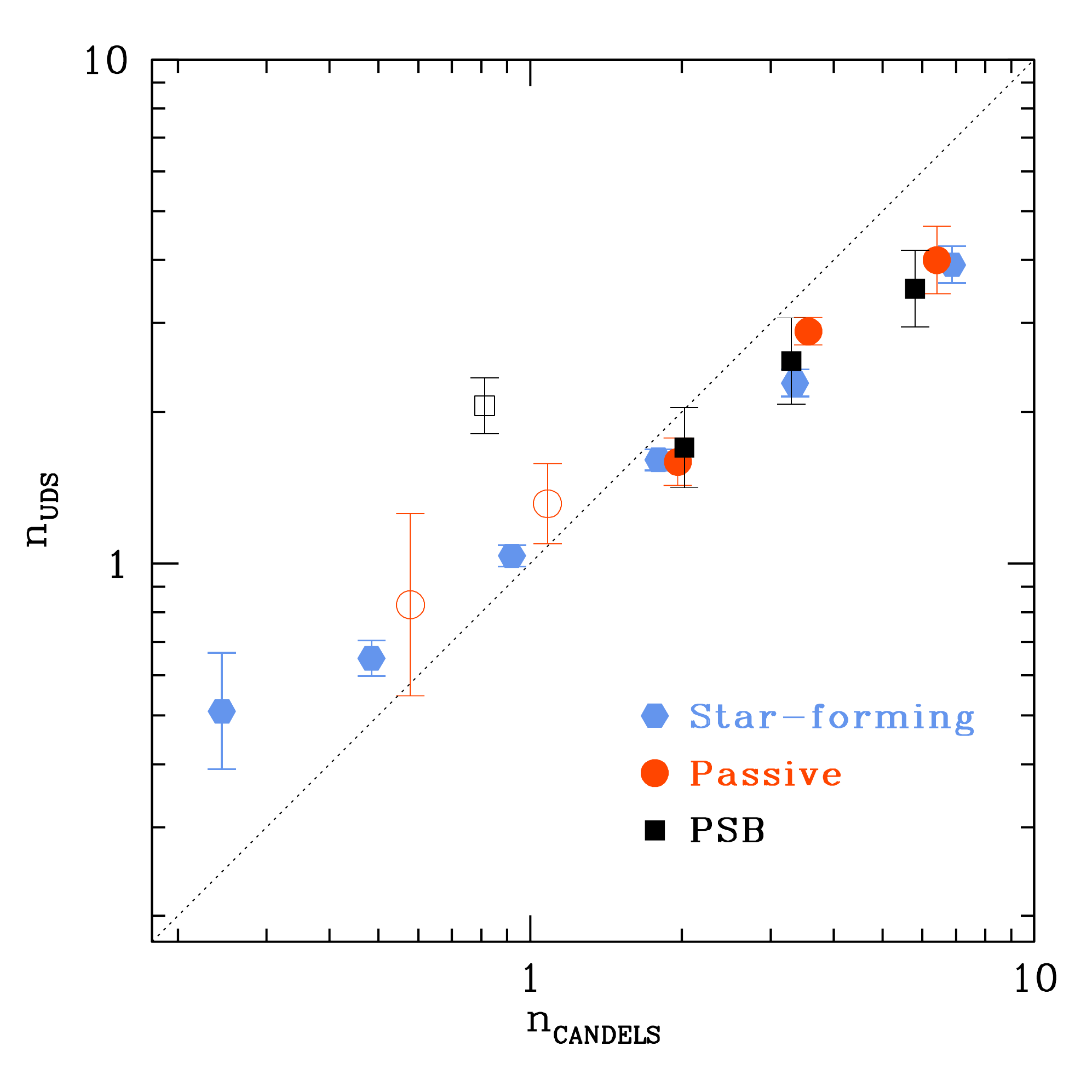}
    \caption{ A comparison of S\'{e}rsic index measurements for
      galaxies measured with ground-based $K$-band imaging and HST
      $H$-band imaging, using the subsample of $\sim$9\% of galaxies
      within the UDS CANDELS mosaic.  The upper panel compares
      individual galaxies, with a characteristic error bar denoting
      the median uncertainty in $\log_{10}(n)$ for the PSB
      population. We find that ground-based determinations of
      S\'{e}rsic indices are sufficient to broadly distinguish
      populations with high average values from those with low average
      values. The lower (binned) distribution compares the mean
      S\'{e}rsic indices, binned as a function of $n_{\rm
        CANDELS}$. Open symbols denote mean values determined from
      fewer than 5 galaxies.  There are deviations from a 1:1
      relation, but no evidence for a systematic bias in ground-based
      S\'{e}rsic determinations with spectral type.  Galaxies shown
      have photometric redshifts in the range $1<z<2$ and stellar
      masses with log (M$_{\ast}$/M$_{\sun}$)$>10$, to match the
      samples used in this work.  }
    \label{fig:sersic-test}
\end{figure}

\subsection{CANDELS-UDS}

Throughout this work we compare our ground-based determinations of
galaxy size and S\'{e}rsic index with measurements from the Hubble
Space Telescope (HST) CANDELS survey (Grogin et al. 2011; Koekemoer et
al. 2011), using measurements provided in van der Wel et al. (2012).
 The UDS is one of the three targets for the CANDELS Wide
survey, with imaging in the $J$ and $H$ bands taken with the Wide
Field Camera 3 (WFC3). The CANDELS imaging covers only $\sim$7\% of
the UDS field ($\sim$9\% of the area used in our analysis), but this
is sufficient to provide an independent test and calibration for our
ground-based structural parameters. Full details of the \textsc{GALFIT}
measurement of structural parameters within CANDELS are given in van
der Wel et al. (2012).

\begin{figure*}
	\includegraphics[width=0.45\linewidth]{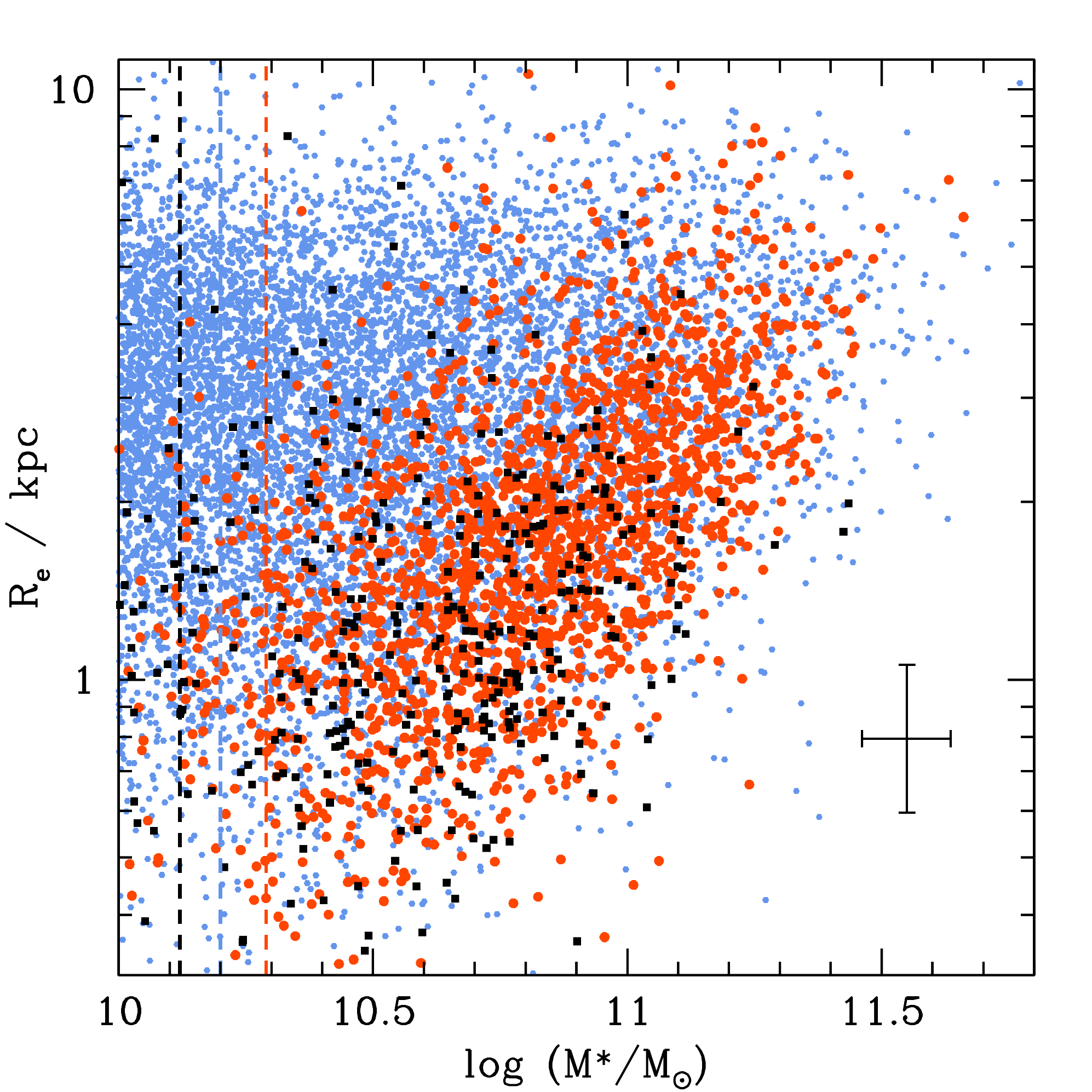}
	\includegraphics[width=0.45\linewidth]{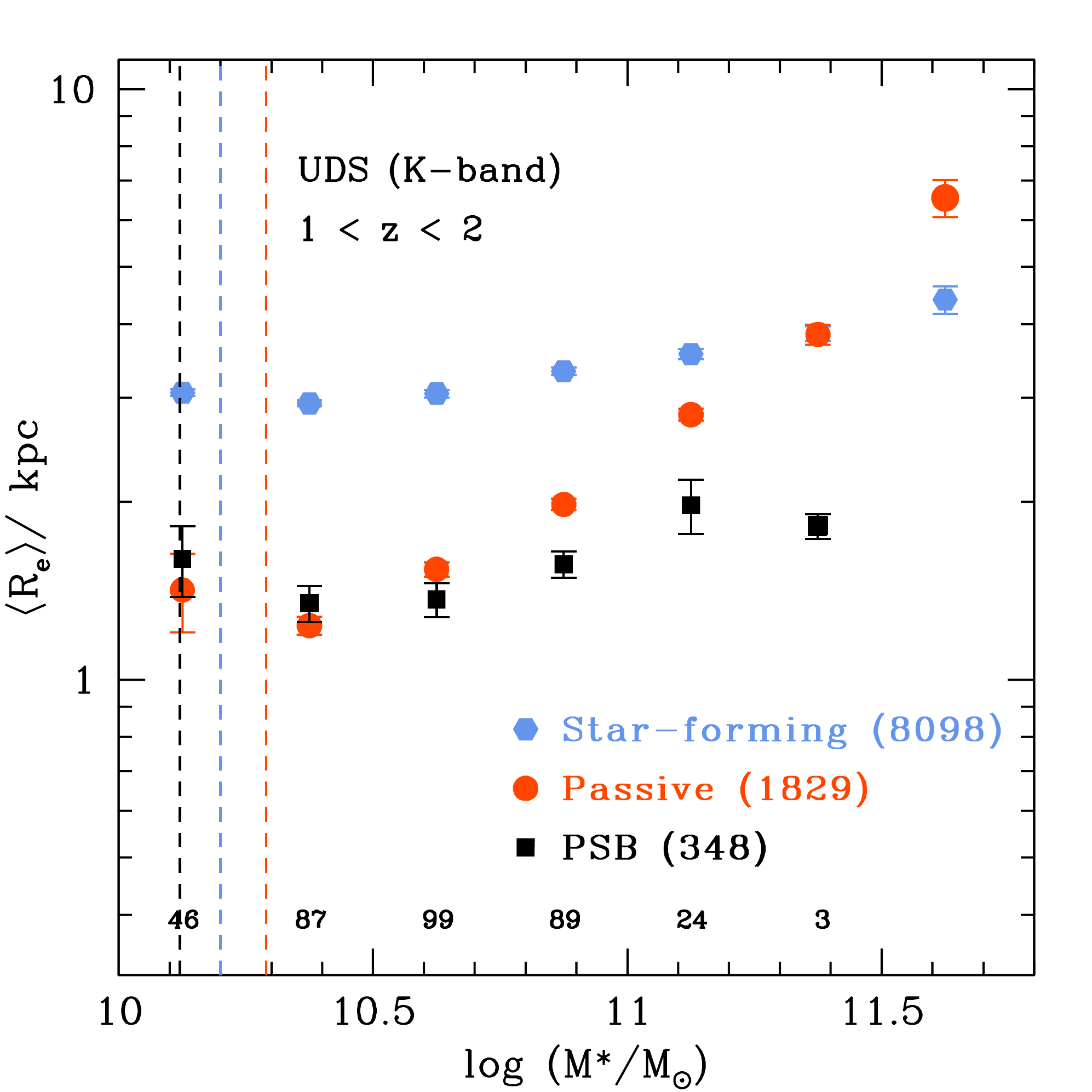}
	\includegraphics[width=0.45\linewidth]{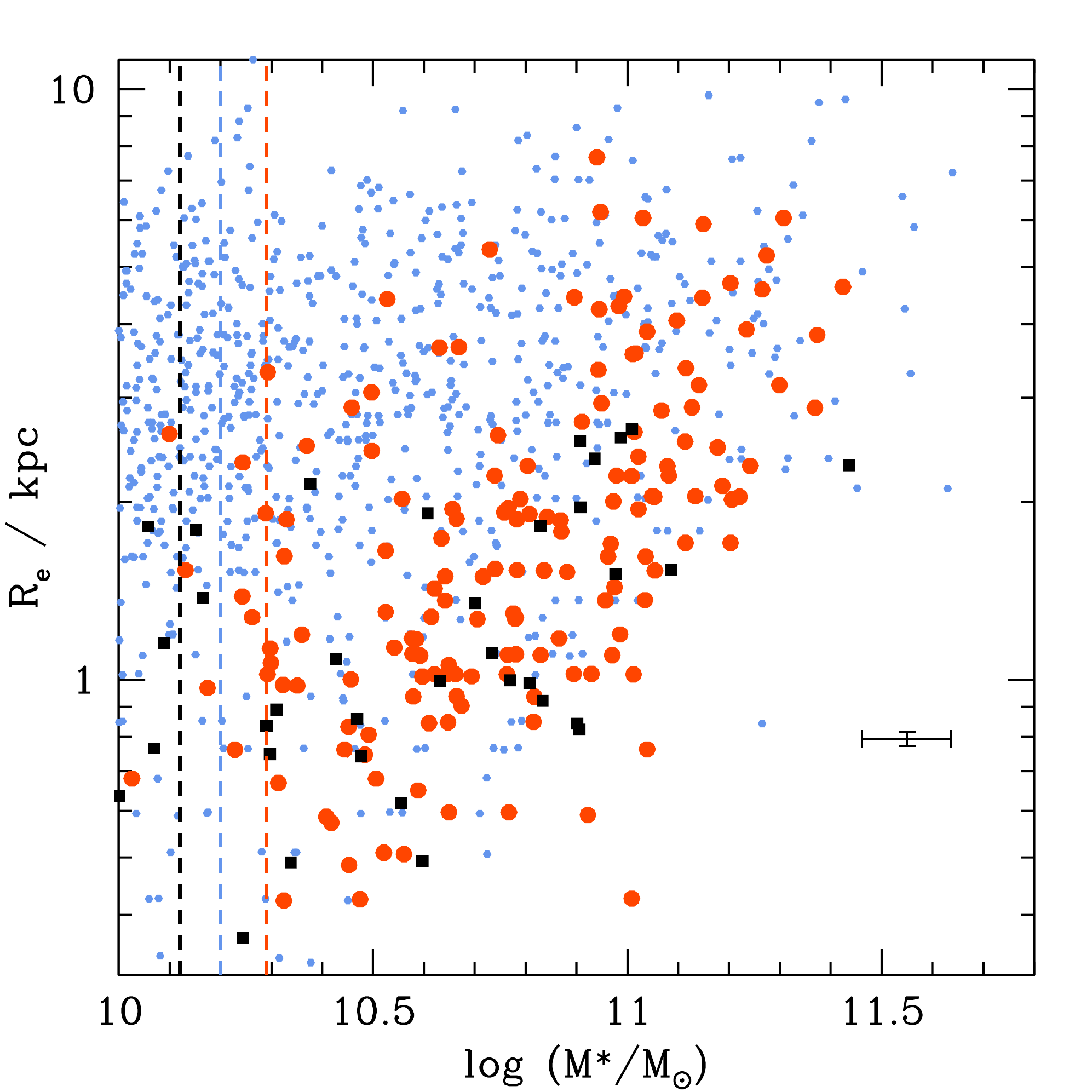}
	\includegraphics[width=0.45\linewidth]{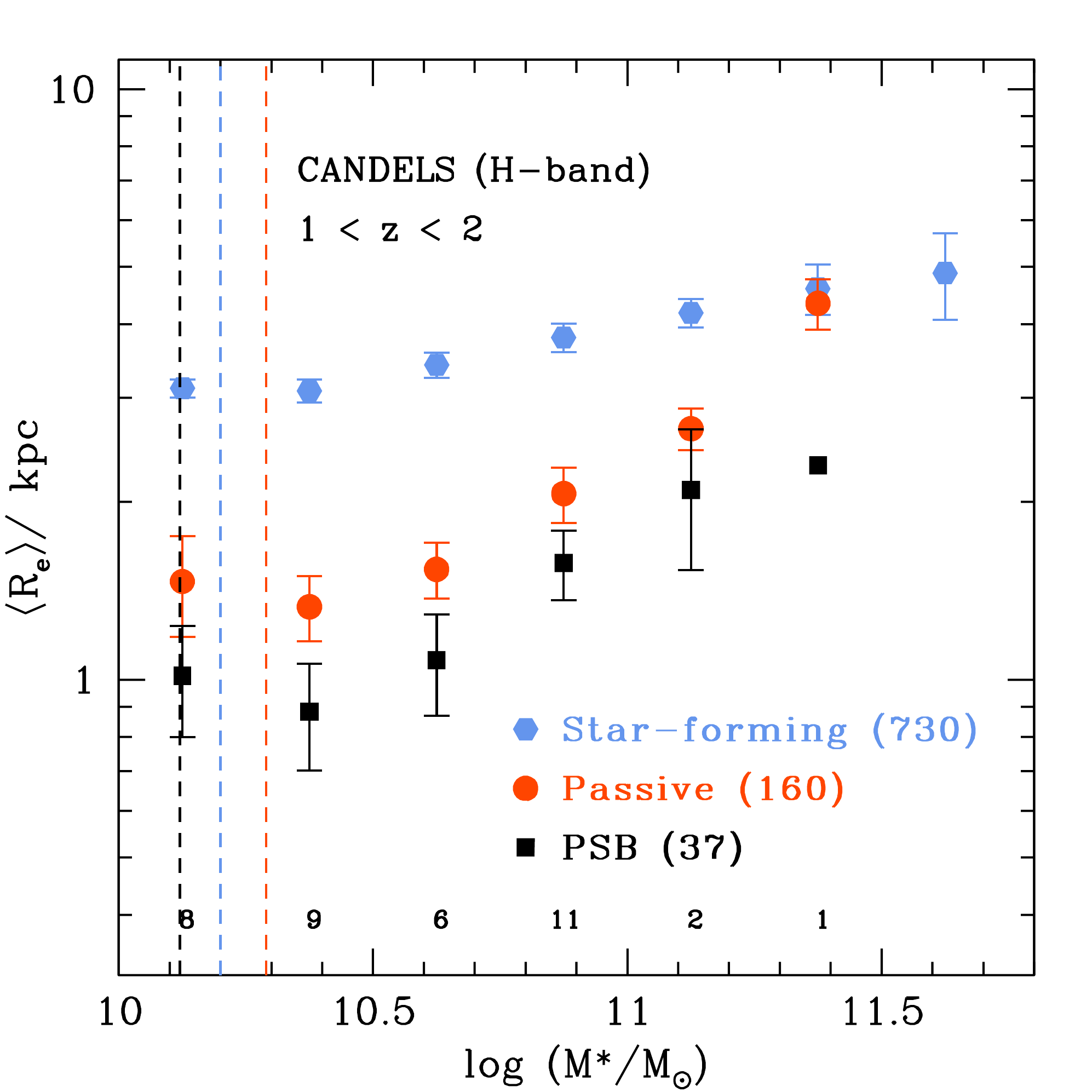}
    \caption{The stellar size-mass relation for star-forming, passive
      and post-starburst galaxies in the redshift range $1<z<2$.  The
      top figures show the results from the ground-based $K$-band
      sample, showing individual galaxies (left) and mean sizes as a
      function of stellar mass (right).  For the individual galaxies
      (left), a characteristic $1\,\sigma$ error bound is displayed,
      showing the median uncertainties in the fitted size and stellar
      mass for the post-starburst population.  For the binned relation
      (right), the numbers in parentheses denote the number of
      galaxies in each sample, while the number of PSBs per mass bin
      is also shown.  Errors on mean sizes represent the standard
      error on the mean (though we caution that the highest mass PSB
      bin contains only 3 galaxies). The lower figures show the
      equivalent for the subsample of galaxies with HST CANDELS
      $H$-band imaging, with sizes determined by van der Wel et
      al. (2012).  Very similar trends are observed, although we note
      the small sample of PSBs (e.g. one galaxy in the highest mass
      bin).  Overall, we conclude that post-starburst galaxies are
      exceptionally compact, with evidence that (on average) they are
      smaller than typical quiescent galaxies at the highest stellar
      masses (M$_{\ast}> 10^{10.5} ~$M$_{\sun}$).  The dashed vertical
      lines denote the 95\% completeness limits, determined at the
      upper redshift range ($z=2$).}
    \label{fig:size-mass}
\end{figure*}

\begin{figure}
        \includegraphics[width=0.9\linewidth]{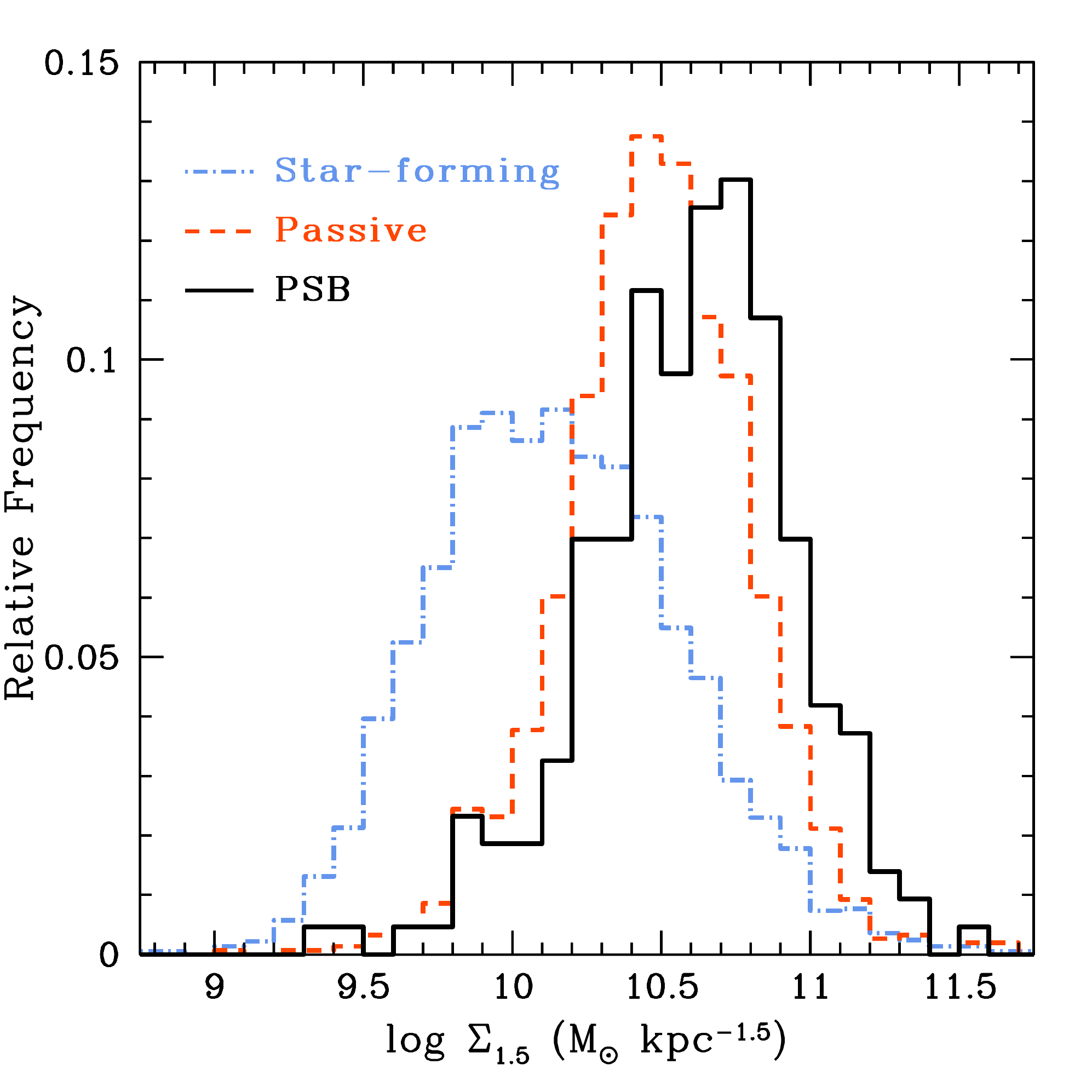}
    \caption{We compare the distribution of $\Sigma_{1.5}$ values for
      the three galaxy populations in the redshift range $1<z<2$, with
      stellar masses M$_{\ast}> 10^{10.5} ~$M$_{\sun}$. This parameter
      ($\Sigma_{1.5} \equiv$M$_{\ast}/R_e^{1.5}$), defined by Barro et
      al. (2013), effectively removes the slope of the galaxy
      mass/size relation; high values of $\Sigma_{1.5}$ correspond to
      galaxies that are compact for their stellar mass. A KS test 
      rejects the null hypothesis that passive galaxies and PSBs are
      drawn from the same underlying distribution in $\Sigma_{1.5}$,
      with a significance of $>99.99\%$.}
    \label{fig:sigma15}
\end{figure}

\begin{figure*}
        \includegraphics[width=0.45\linewidth]{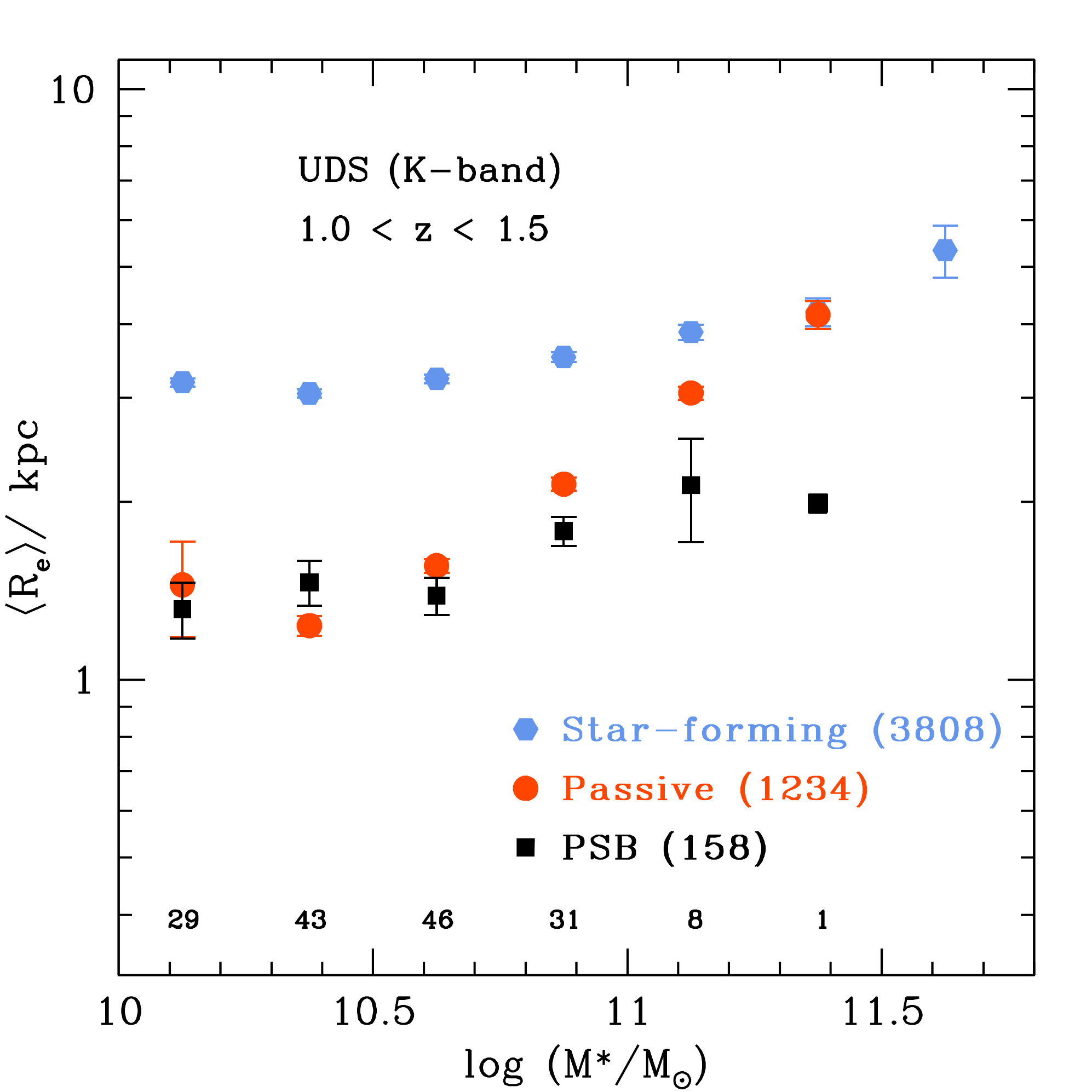}
        \includegraphics[width=0.45\linewidth]{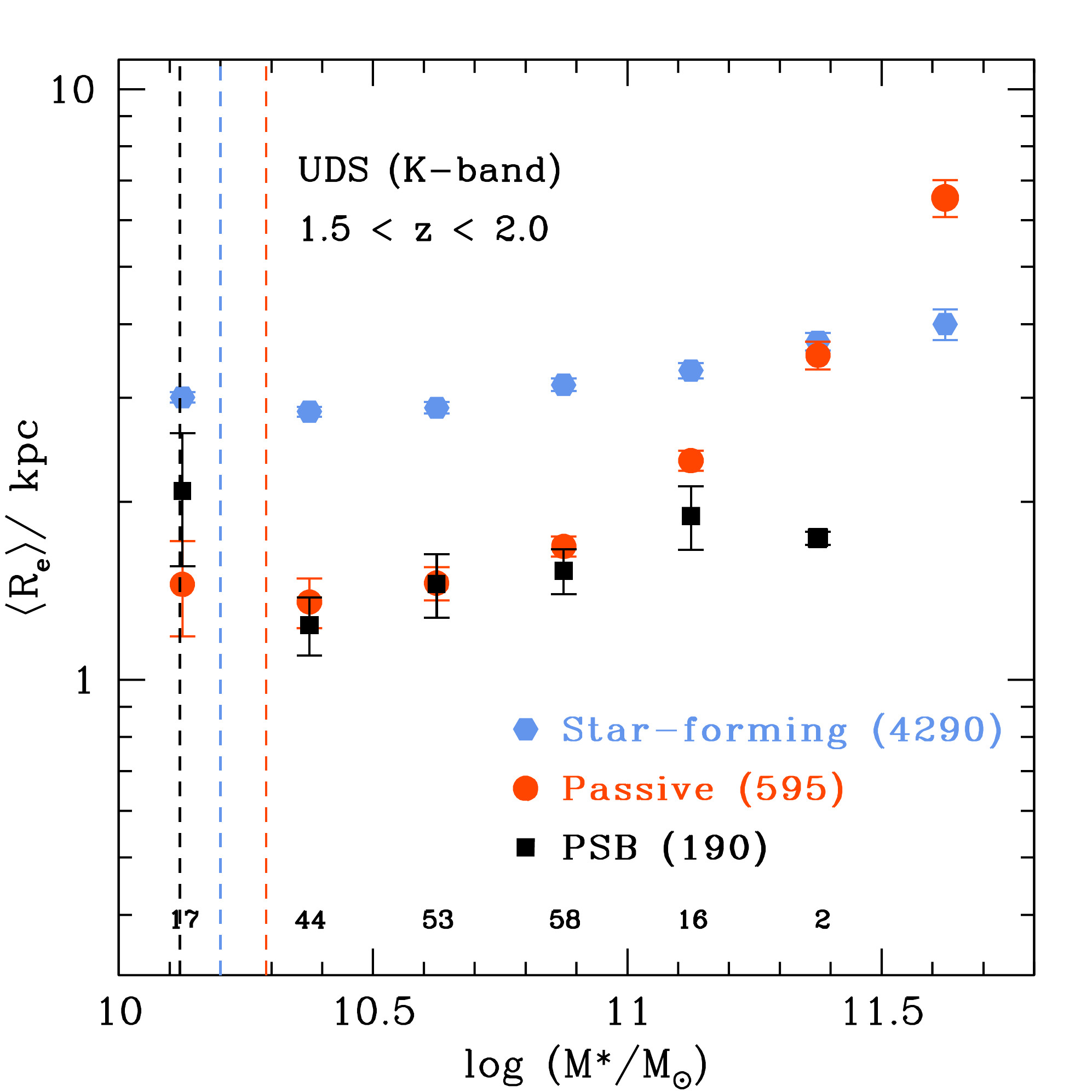}
    \caption{A comparison of the galaxy mean size--stellar mass
      relation for star-forming, passive and post-starburst galaxies
      in two redshift bins.  The numbers in parentheses denote the
      number of galaxies in each sample, while the number of PSBs per
      mass bin is shown at the bottom of each plot.  Errors on mean
      sizes represent the standard error on the mean (though we note
      the small number of PSBs in the highest mass bins).  The
      vertical lines denote the 95\% completeness limits (all below
      $10^{10} ~$M$_{\sun}$ for the low-redshift bin). }
    \label{fig:size-mass-z12}
\end{figure*}

\section{Ground-based measurements of size and S\'{e}rsic index}
\label{sec:sizesersic}

We determined structural parameters for the $K$-band galaxy sample
using the \textsc{GALAPAGOS} software (Barden et al. 2012).  The
package allows the automated use of \textsc{GALFIT} (Peng et al. 2002)
to fit S\'{e}rsic light profiles (S\'{e}rsic 1968) to all galaxies in
the UDS, parameterized with a S\'{e}rsic index, $n$, and effective
radius, $R_e$, measured along the semi-major axis.  We acknowledge
that many high-redshift galaxies are described by more complex
morphologies (see Bruce et al. 2014), but single S\'{e}rsic fits
provide a simple parameterization to allow us to compare the bulk
properties of the galaxy populations.

An accurate determination of the point-spread function (PSF) is
critical for this process, as galaxies at $z>1$ typically have
half-light radii below 0.5 arcsec. Following the work of Lani et
al. (2013), we investigated PSF variations across the UDS field and
found that most variation occurred between WFCAM detector boundaries
within the UDS mosaic (Casali et al. 2007).  Testing revealed that we
could obtain consistent results by splitting the UDS field into 16
overlapping sub-regions, corresponding to the $4\times 4$ WFCAM tiling
pattern. Within each region the light profiles of approximately 100
stars were stacked to provide the local PSF measurement, with
variations across the field in the range $0.75-0.81$ arcsec
(FWHM). Considerable care was taken to mask sources in the vicinity of
the stars used for PSF measurement.

From the resulting measurements of size and S\'{e}rsic index we rejected
$\sim$9\% of galaxies where GALFIT failed to converge on a S\'{e}rsic
solution.  The rejection rate was similar for the star-forming,
passive and post-starburst populations. A further $\sim$1\% of
galaxies were rejected (a-priori) if the fits were formally very poor
($\chi^2_\nu>100$), which typically corresponded to highly blended
objects on the $K$-band image. Matching the output from
\textsc{GALAPAGOS} with our Supercolour catalogue, we obtain a final
sample of 8098 star-forming galaxies, 1829 passive galaxies, and 348
PSBs in the redshift range $1<z<2$ with M$_{\ast}> 10^{10}
~$M$_{\sun}$.

In Figures \ref{fig:size-test} and \ref{fig:sersic-test} we display
the resulting size and S\'{e}rsic measurements for the subset of UDS
galaxies within the HST CANDELS survey. Ground-based measurements are
compared with those obtained using H-band CANDELS measurements, as
published in van der Wel et al. (2012).  

For the size measurements, we find a  tight relationship between
the ground-based $K$-band and CANDELS $H$-band sizes, as previously
found by Lani et al. (2013).  On average, the sizes obtained from
CANDELS are systematically $\sim$13\% larger, which is consistent
with previous comparisons of size measurements as a function of
waveband (e.g. Kelvin et al. 2012). The systematic offset is
consistent among the three galaxy populations studied here so we apply
no corrections for this effect.  The characteristic scatter in $\delta
R/R$, given by the normalized median absolute deviation ($\sigma_{{\rm
    NMAD}}$) is 17\%, 16\% and 24\% for the star-forming, passive and
PSB populations respectively.

Ground-based measurements of S\'{e}rsic index are far more uncertain
for a given galaxy (Figure \ref{fig:sersic-test}), and we find a
significant degree of scatter when comparing ground-based and CANDELS
measurements.  Formally, the characteristic scatter in $\delta n/n$,
given by the normalized median absolute deviation ($\sigma_{{\rm
    NMAD}}$) is 45\%, 39\%, and 35\% for the star-forming, passive and
PSB populations respectively.  Nevertheless, there is a clear
correlation, and we find that ground-based determinations are
sufficient to distinguish populations with ``high'' S\'{e}rsic indices
(e.g. $n>2$) from those with ``low'' S\'{e}rsic indices
($n<2$). Comparing the three primary galaxy types, we find consistent
results; the passive and post-starburst populations show consistently
high S\'{e}rsic indices (from ground or spaced-based measurements),
while star-forming galaxies are concentrated at lower values.  The binned
distribution demonstrates that the correlation between ground and
space-based S\'{e}rsic indices is not perfectly 1:1, but we see no
systematic differences in this relation between the three populations.
We conclude that ground-based measurements can be used to broadly
compare the S\'{e}rsic indices for our galaxy populations. In
addition, we will use the subset of galaxies with HST measurements
($\sim$9\%) to verify any conclusions drawn from the larger
ground-based sample.  The S\'{e}rsic distributions will be compared
further (and as a function of stellar mass) in Section
\ref{sec:sersic}.

\section{The sizes of post-starburst galaxies}
\subsection{The size--mass relation}
\label{sec:sizemass}

In Figure \ref{fig:size-mass} we compare the size versus stellar-mass
relation for galaxies in the redshift range $1<z<2$. Individual
galaxies are shown, along with mean values as a function of stellar
mass (in bins of 0.25 dex).  In the upper panels we show the results
from the ground-based $K$-band imaging, while the lower panels are
based on independent sizes from CANDELS $H$-band imaging (covering
$\sim$9\% of the sample).  The 95\% mass completeness limits are shown
(see Section \ref{sec:masses}), determined at $z=2$ to provide
conservative limits.  In determining mean values for the ground-based
sample we applied a 5$\sigma$ clip (with one iteration), to remove
extreme outliers, but removing this constraint has no significant
influence.

Representative error bounds for individual galaxies are shown on the
left panels, based on the median errors on the PSB sample. For the
CANDELS sizes, data are taken from van der Wel et al. (2012), which
include estimates for random and systematic errors from GALFIT. For
the ground-based individual errors, we add in quadrature the scatter
in $\delta R/R$ determined by the comparison with CANDELS (Section
\ref{sec:sizesersic}).  The representative uncertainty on stellar
masses is based on our Bayesian mass-fitting analysis described in
Wild et al. (2016), as briefly outlined in Section \ref{sec:masses}.

The size--mass relations show the expected trends for star-forming and
passive galaxies, consistent with previous studies (e.g. Daddi et
al. 2005; Trujillo et al. 2006; van Dokkum et al. 2008; van der Wel et
al. 2014; McLure et al. 2013).  On average, passive galaxies appear
significantly more compact than star-forming galaxies of equivalent
stellar mass, but show a steeper size--mass relation, leading to
convergence at the highest masses (M$_{\ast} \sim 10^{11.5}
~$M$_{\sun}$). Intriguingly, we find that post-starburst galaxies at
this epoch are also extremely compact; they are comparable in size to
the established passive galaxies, with evidence that they are smaller
on average at high mass (M$_{\ast}> 10^{10.5} ~$M$_{\sun}$).  These
trends are apparent with the large ground-based sample and with the
smaller CANDELS sample.

We performed a bootstrap analysis as a simple test of significance,
randomly sampling the (ground-based) populations within each mass bin,
with replacement.  For the 4 bins above $10^{10.5} ~$M$_{\sun}$, the
fraction of the resampled populations in which the passive galaxies
show mean sizes equal to (or smaller than) the mean of the PSB
population are ~$0.07$, $5\times10^{-5}$, $1.2\times 10^{-4}$,
$4\times10^{-5}$ (low to high mass, respectively).  We note that the
final bin contains 86 passive galaxies, but only 3 PSBs, so the
bootstrap comparison for this bin may be unreliable.  Overall,
assuming no systematic errors, we find evidence that massive
post-starburst galaxies at $z>1$ are significantly more compact, on
average, than passive galaxies of comparable mass.  Repeating the
analysis using the median sizes produced very similar trends. The
median analysis and further tests of robustness are presented in
Appendix A.  Our results are consistent with the findings of Yano et
al. (2016), who also found evidence that high-redshift post-starburst
galaxies are very compact.

As an additional comparison, in Figure \ref{fig:sigma15} we display
the distribution of $\Sigma_{1.5}\equiv M_{\ast}/R_e^{1.5}$ for the
three galaxy populations, measured for the redshift range $1<z<2$ and
stellar masses M$_{\ast}> 10^{10.5} ~$M$_{\sun}$.  Following Barro et
al. (2013), we use this parameter to effectively remove the slope in
the galaxy size/mass relation.  Fitting a function of the form
M$_{\ast} = \Sigma ~ R_e^{\alpha}$ to the passive population, we find
a best fit with $\alpha=1.55$, in very good agreement with the value
$\alpha=1.5$ assumed by Barro et al. (2013).  A simple
Kolmogorov-Smirnov (KS) test rejects the null hypothesis that passive
galaxies and PSBs are drawn from the same underlying distribution in
$\Sigma_{1.5}$ with a significance $>99.99$\%, with the same significance
obtained with either value of $\alpha$.

Given the strong evolution of the PSB mass function (Wild et
al. 2016), a concern is that massive PSBs are more common at higher
redshifts, which may bias the size--mass comparison when measured over
a wide redshift range. In Figure \ref{fig:size-mass-z12} we therefore
display the size--mass relation in two narrower redshift bins,
$1.0<z<1.5$ and $1.5<z<2.0$. With two independent samples, the results
confirm that post-starburst galaxies, on average, show smaller
half-light radii than the passive population at high mass
(M$_{\ast}> 10^{10.5} ~$M$_{\sun}$).  An additional test was performed
using a weighted mean, with a redshift-dependent weight determined for
each PSB using the ratio of the $n(z)$ distributions for passive
galaxies and PSBs. The resulting size--mass relations were barely
changed, with only a slight reduction in the significance of the
differences reported above.

A natural interpretation of our findings is that quiescent galaxies
are most compact when they are newly-quenched, but then grow with
cosmic time. Given the short-lived nature of the PSB phase, we expect
the majority of  passive galaxies to have gone through a similar
stage in their past (Wild et al. 2016).  Our results therefore provide
evidence for the genuine growth of individual galaxies, suggesting
that the growth of the population as a whole is not purely caused by a
progenitor bias. We discuss the implications further in Section \ref{sec:discussion}.

A simple calculation allows us to compare the size differences with the
observed cosmological growth.  Based on population synthesis models,
we estimate that the established passive population quenched
approximately 0.5--1 Gyr before the PSB population at these redshifts
(Wild et al. 2016), and the characteristic difference in size
(e.g. based on the shift in $\Sigma_{1.5}$) is approximately 25\% on
average at M$_{\ast}> 10^{10.5} ~$M$_{\sun}$. The implied growth rate
is similar  to the observed {\em cosmological}
growth, parameterized by van Dokkum et al. (2010) in the form $r_e
\propto (1+z)^{-1.3}$ (i.e. $\sim$25\% per Gyr at $z=1.5-2$). An
improvement on these tentative conclusions will require more accurate
age-dating of the stellar populations, which will soon be possible
with growing spectroscopic samples.

\begin{figure*}
	\includegraphics[width=0.45\linewidth]{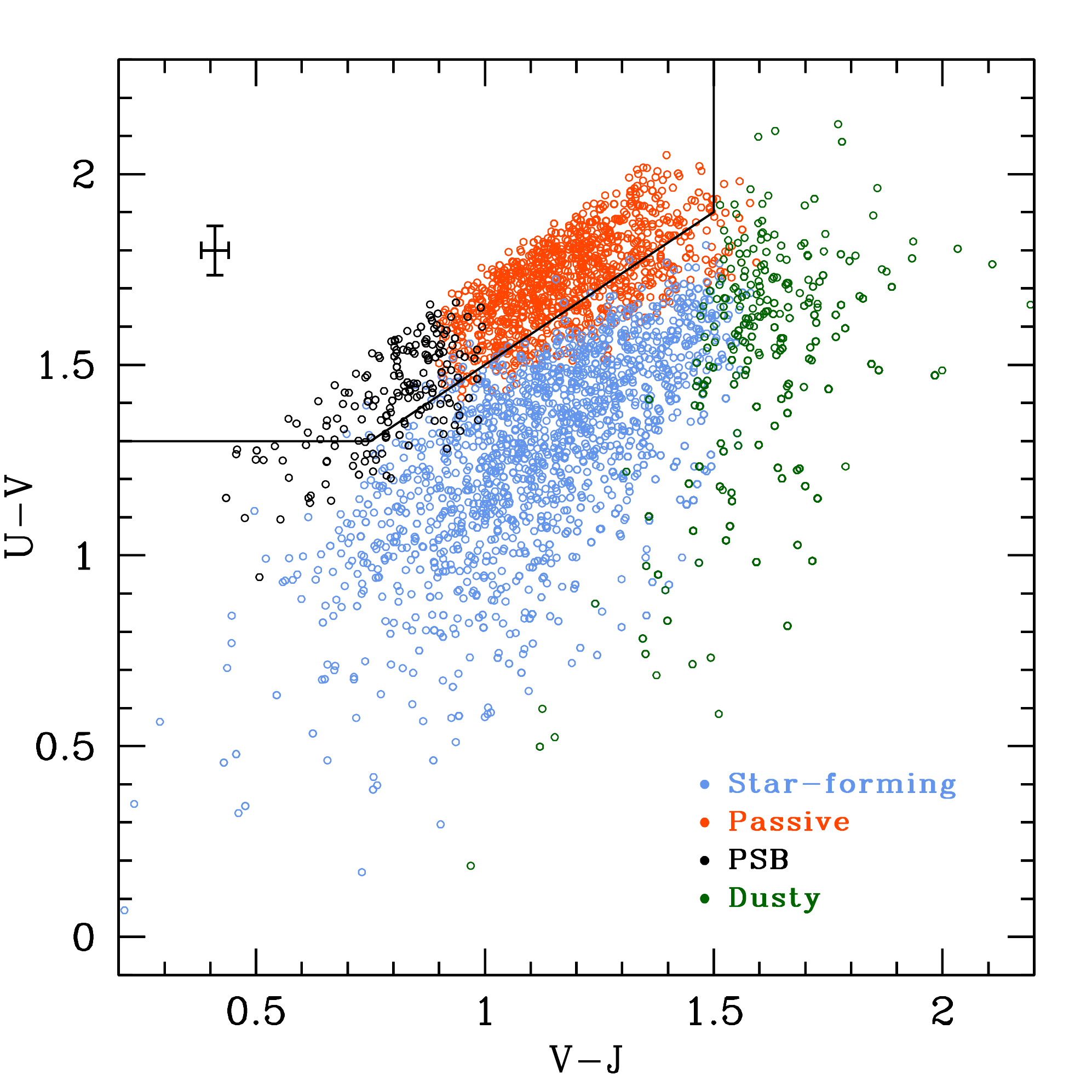}
	\includegraphics[width=0.45\linewidth]{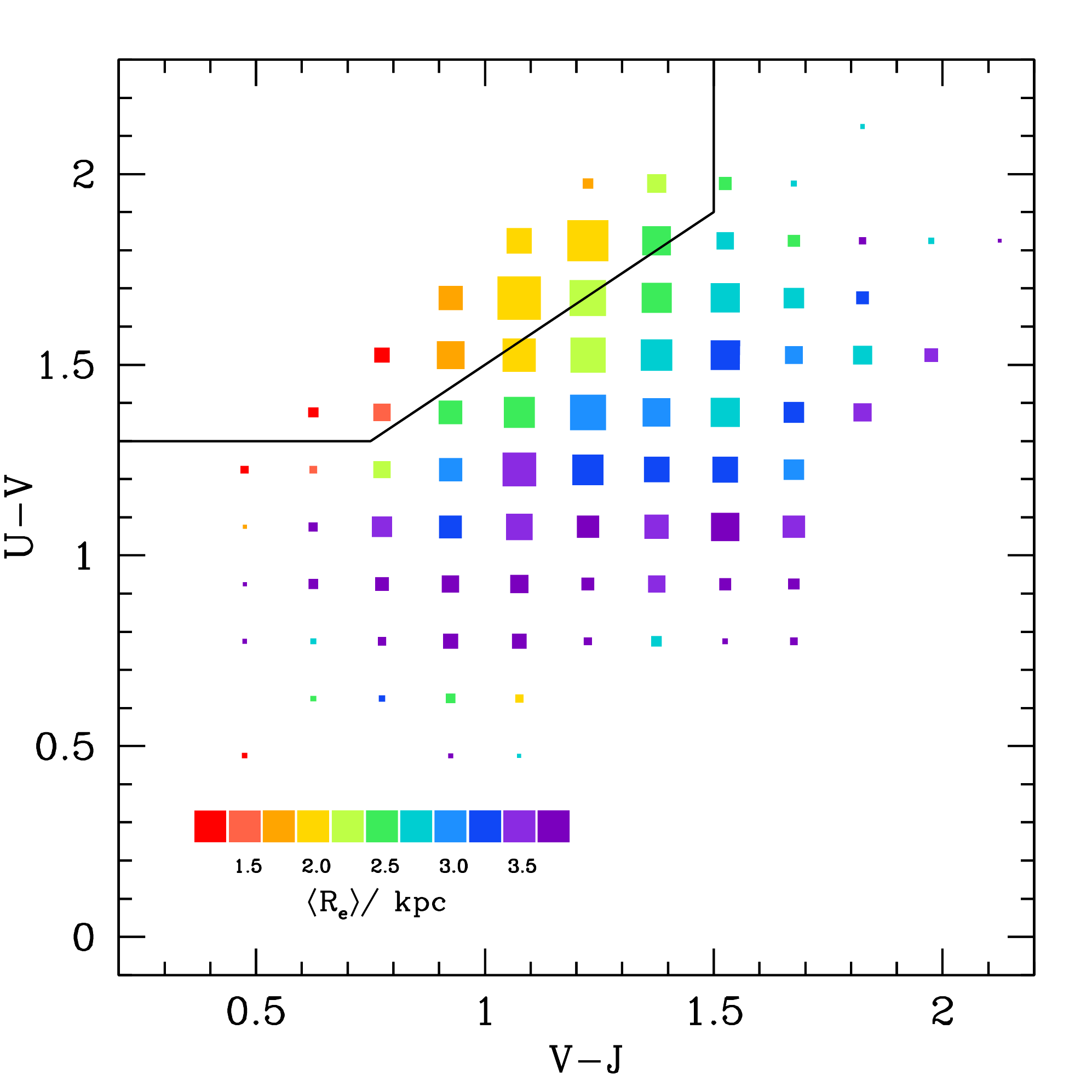}
    \caption{Rest-frame $UVJ$ colour-colour diagrams for UDS galaxies
      in the redshift range $1<z<2$, and within the mass range $10.5<
      \log ~($M$_{*}/$M$_{\sun})<11.5$. Previous studies have
      separated galaxies into ``quiescent'' and ``star-forming''
      categories using the boundary shown. The left panel compares the
      classification of galaxies using supercolours with the UVJ
      method (in this diagram separating the dusty PCA class from
      the other star-forming galaxies). We find good agreement between the classification methods, with a very similar
      boundary between star-forming and passive types.  Post-starburst
      galaxies selected by supercolours are primarily located at the
      blue end of the quiescent UVJ region, in good agreement with
      the boundaries proposed by Whitaker et al. (2012). A
      representative error bar is displayed, showing the median
      uncertainties for the PSB population.  The right-hand panel
      shows average sizes determined for all galaxies in colour-colour
      bins, without using supercolours. Symbol size is proportional to
      the number of galaxies per bin.  The trends confirm that
      ``young-quiescent'' galaxies selected by the $UVJ$ technique
      show smaller average sizes than redder passive galaxies,
      consistent with our classification based on PCA supercolours.}
    \label{fig:uvjfig}
\end{figure*}

\subsection{Stellar mass uncertainties}
\label{sec:uncertainties}

In this section we explore the potential impact of both random and
systematic errors on our stellar masses. 

The typical uncertainty from our Bayesian stellar mass fitting is
$\sigma \simeq$ 0.1 dex  for all galaxy types,
allowing for the degeneracy between fitted parameters and the
uncertainties on photometric redshifts. To investigate the impact on
our conclusions, we performed Monte Carlo realisations, allowing the
stellar masses to shift randomly within a Gaussian probability
distribution in log M$_{\ast}$. We found no impact on any of the
results presented in Section \ref{sec:sizemass}.  Comparing the
resulting distributions in $\Sigma_{1.5}$, the significance of the
difference between the PSB and passive galaxy populations was
unchanged, suggesting that the sample overall is large enough to
minimise the influence of random errors.

To investigate the influence of fitting methods, we re-evaluated the
size--mass relations using two independent sets of stellar masses
derived by Simpson et al. (2013) and Hartley et al. (2013), the latter
also using an independent set of photometric redshifts.  No
significant differences were found. 

As a note of caution, however, we acknowledge that the observed
differences between PSBs and passive galaxies could arise if our
stellar masses are {\em systematically} overestimated for younger
stellar populations. Our stellar masses are based on population
synthesis models from BC03, which may underestimate the influence of
thermally-pulsing asymptotic giant branch (TP-AGB) stars (Maraston et
al. 2006). Such stars may have a major contribution to the rest-frame
near-infrared light for stellar populations in the age range 0.2 to 2
Gyr, potentially leading to overestimated stellar masses for passive
galaxy populations.  The influence of TP-AGB stars is discussed
further in Wild et al. (2016), where it was found that the influence
of TP-AGB stars is strongest for galaxies with BC03-determined ages
$>1$~Gyr, for which the ages and mass-to-light ratios are reduced
using the models of Maraston et al. (2006). The net effect would be
to move our passive galaxies to lower masses relative to the younger
PSB population, which would {\em enhance} the differences in the
size--mass relation.

In summary, our findings appear robust to known sources of random and
systematic error, but we acknowledge the possibility that unknown
systematic uncertainties in the stellar mass determination may
contribute to the difference in size--mass relations presented in our
work. Future deep infrared spectroscopy may allow a more detailed
investigation of the inherent uncertainties in determining stellar
masses from photometric data.

\subsection{A comparison with UVJ selection}
\label{sec:uvj}

To allow a comparison of our supercolour technique with previous work,
in Figure \ref{fig:uvjfig} (left) we present a rest-frame UVJ
colour-colour diagram for UDS galaxies in the redshift range
$1<z<2$. As previously shown in W14, the classification of galaxies
using supercolours agrees very well with the more traditional UVJ
colour selection (Labb\'e et al. 2005; Wuyts et
al. 2007). Post-starburst galaxies are generally found at the blue end
of the passive UVJ region, in good agreement with the findings of
Whitaker et al. (2012). The distribution in $U-V$ alone confirms that
PSBs predominantly lie in the classic ``green valley'', intermediate
between passive and star-forming galaxies, but the addition of the
$V-J$ colour isolates this population in the distinct region
corresponding to the youngest red-sequence galaxies.

In the right-hand panel of Figure \ref{fig:uvjfig} we illustrate the
average effective radii in colour-colour bins.  We select the subset
of galaxies over the mass range $10.5< \log
~($M$_{*}/$M$_{\sun})<11.5$ to minimise the effects of the size--mass
relation.  
The trends
confirm that ``young quiescent'' galaxies selected by the UVJ
technique show smaller average sizes, in good agreement with our
analysis based on supercolours. 
Our results are  consistent with
a similar recent analysis by Yano et al. (2016).

\section{The S\'{e}rsic indices of post-starburst galaxies}
\label{sec:sersic}

\begin{figure*}
	\includegraphics[width=0.32\linewidth]{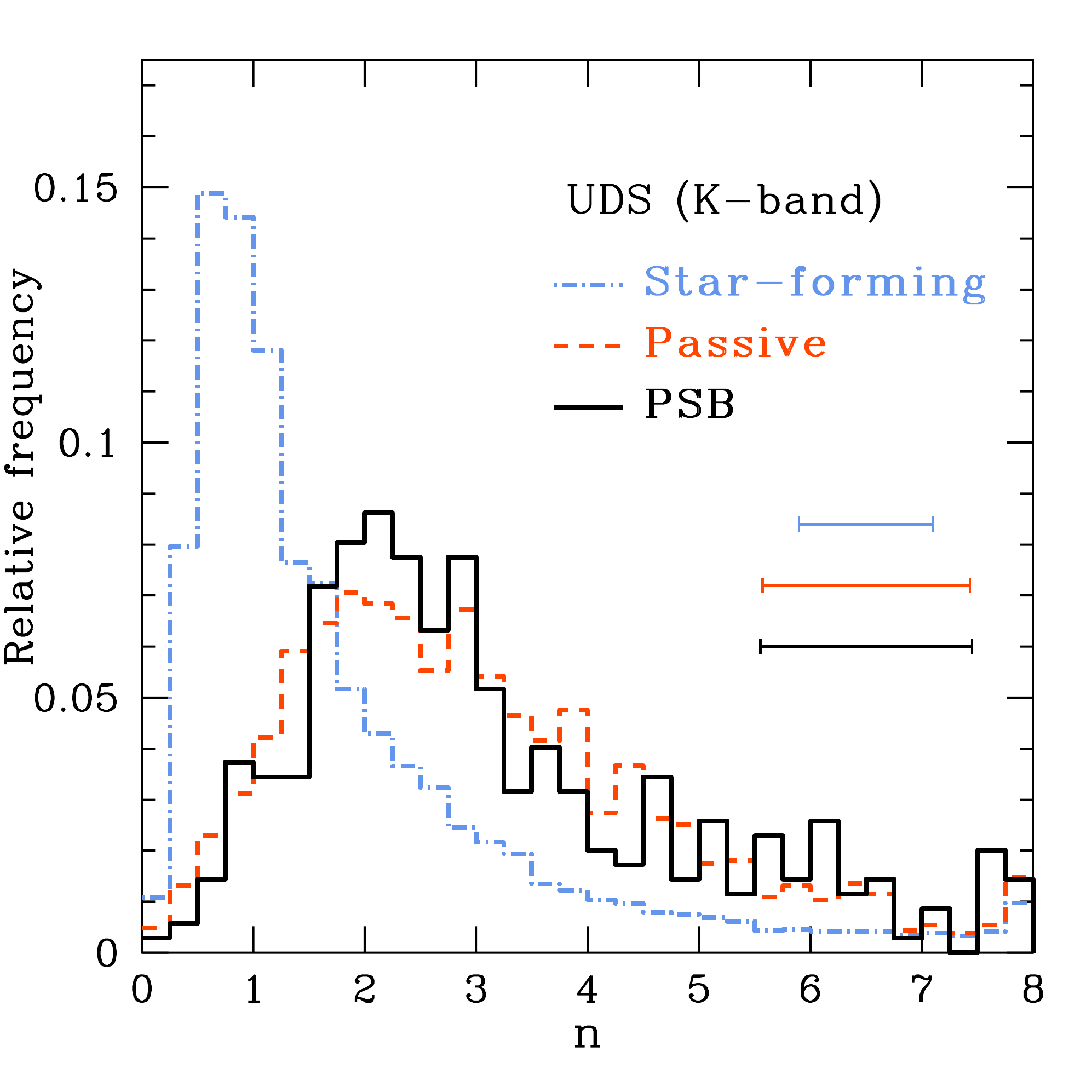}
	\includegraphics[width=0.32\linewidth]{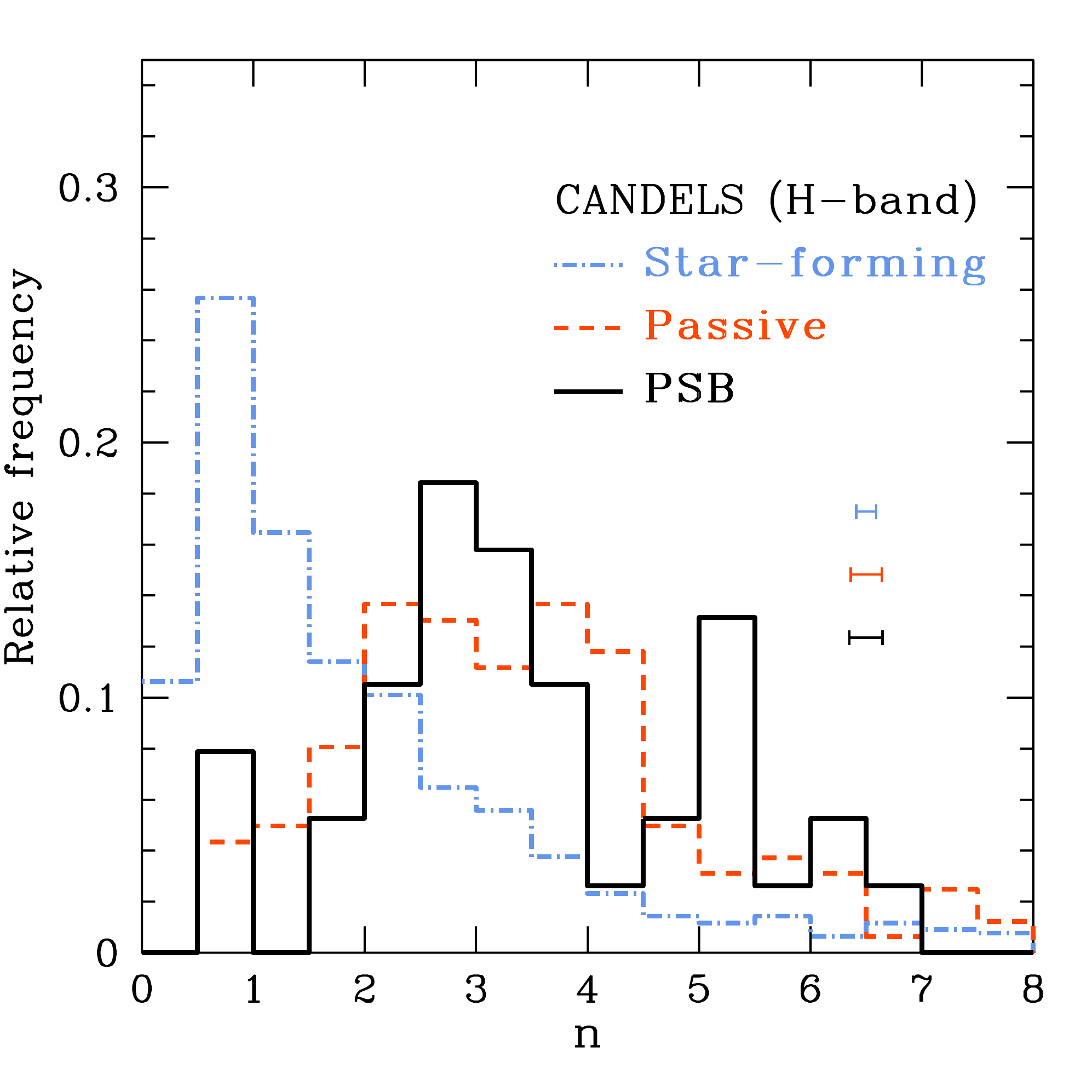}
	\includegraphics[width=0.32\linewidth]{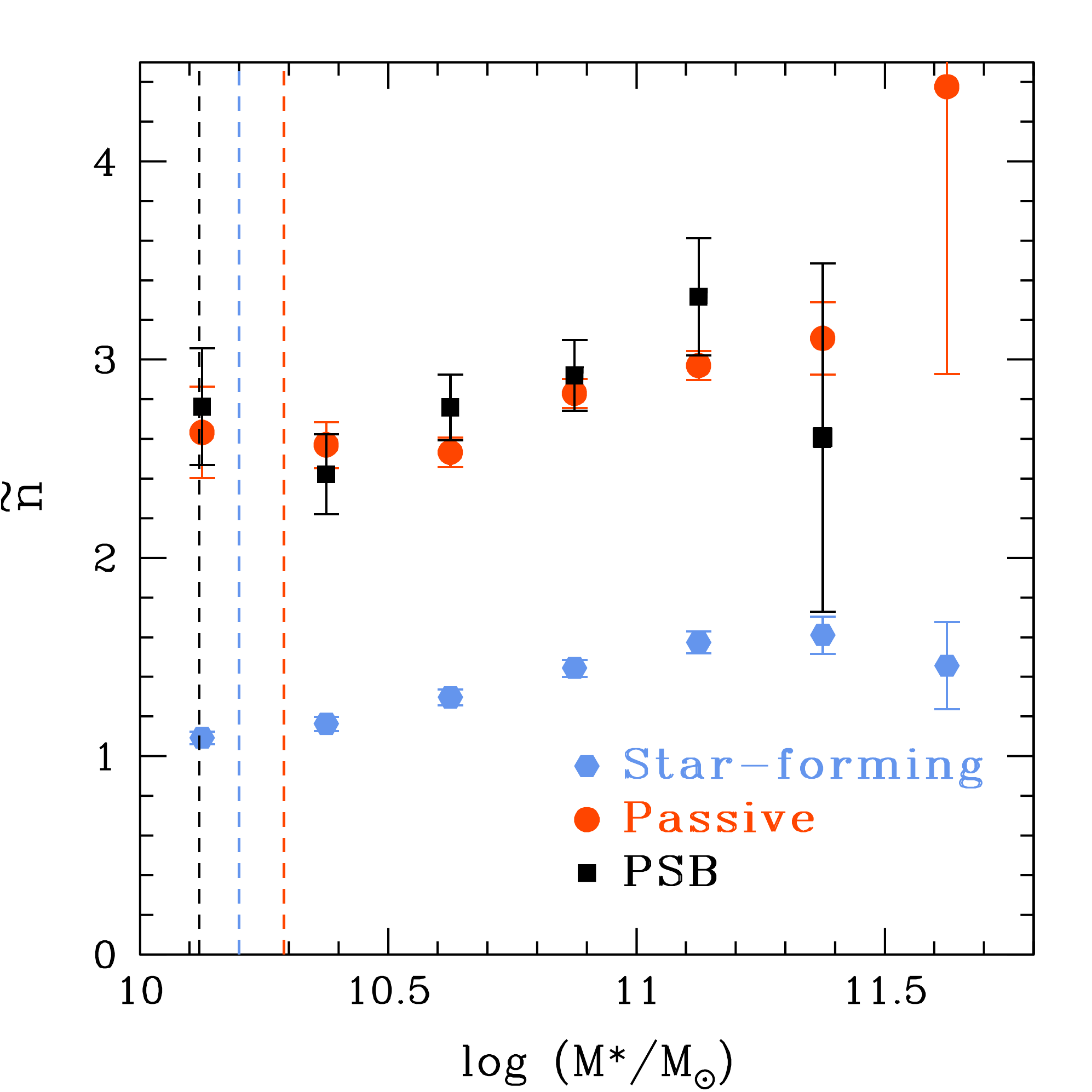}
    \caption{ A comparison of the S\'{e}rsic indices for star-forming,
      passive and post-starburst galaxies in the redshift range
      $1<z<2$ with stellar masses M$_{*}> 10^{10}~$M$_{\sun}$. The
      left panel shows a histogram of S\'{e}rsic indices determined
      from ground-based UDS $K$-band imaging.  The central panel shows
      a histogram of S\'{e}rsic indices determined for the smaller
      sample with available HST $H$-band imaging from CANDELS (from
      van der Wel et al. 2012).  Characteristic uncertainties are
      displayed for each population, showing the median error
      on the fitted S\'{e}rsic index for individual galaxies.  The
      right panel shows median S\'{e}rsic indices as a function of
      stellar mass for the larger $K$-band sample, using identical
      binning to Figure \ref{fig:size-mass}.  We conclude that
      post-starburst galaxies show significantly higher S\'{e}rsic
      indices than star-forming galaxies, but with a distribution that
      is indistinguishable from the older passive population.  }
    \label{fig:sersicfig}
\end{figure*}

In Figure \ref{fig:sersicfig} we compare the S\'{e}rsic indices for
star-forming, passive and post-starburst galaxies at $z>1$. The
distributions obtained from the ground-based $K$-band data are
consistent with those obtained for the smaller CANDELS sample. In both
cases, we find that star-forming galaxies show a distribution peaking
sharply at $n\simeq 1$, while passive and post-starburst galaxies show
very different distributions, peaking at significantly higher values.
A Kolmogorov-Smirnov test confirms these findings, rejecting the null
hypothesis that either passive or post-starburst galaxies are drawn
from the same distribution as star-forming galaxies to a high level of
significance ($>>99.99$\%, using the ground-based sample). In
contrast, the S\'{e}rsic distributions for passive and post-starburst
galaxies do not appear significantly different.

The right-hand panel in Figure \ref{fig:sersicfig} presents the median
S\'{e}rsic indices as a function of stellar mass. We find 
evidence for a slight increase in the median S\'{e}rsic index with
stellar mass for all populations, but in all mass bins the
post-starburst galaxies show significantly higher S\'{e}rsic indices
than star-forming galaxies, and values consistent 
with the passive population. Using the mean produced very
similar trends, but S\'{e}rsic indices were slightly higher in all
cases (by $\delta n\simeq 0.5$).

We caution that single S\'{e}rsic fits provide only a crude
parameterisation of the data, as it is now well-established that most
galaxies at this epoch have more complex morphologies (e.g. Bruce et
al. 2014). A high S\'{e}rsic index does not necessarily imply a purely
spheroidal system, and low S\'{e}rsic indices do not necessarily imply
the presence of an established disc (Mortlock et
al. 2013). Nevertheless, it is clear from our data that the
post-starburst galaxies are structurally very different to the
actively star-forming population, and more comparable to ultra-compact
equivalents of the passive population. We explore multiple-component
fitting in future work (Maltby et al., in preparation).

\section{Discussion}
\label{sec:discussion}

We have presented evidence that massive (M$_{\ast}> 10^{10.5}
~$M$_{\sun}$) recently-quenched (post-starburst) galaxies at high
redshift ($z>1$) are exceptionally compact.  Furthermore, they show
high S\'{e}rsic indices, indistinguishable from the established
passive population at the same epoch. We conclude that the structural
transformation must have occurred before (or during) the event that quenched
their star formation.  Given that the majority of massive passive
galaxies at $z>1$ are thought to have passed through a post-starburst
phase (Wild et al. 2016), our findings suggest a strong link between
quenching and the formation of a compact spheroid.

Our results confirm the findings of Whitaker et al. (2012), who found
evidence that younger passive galaxies are more compact at $z>1$, and
the more recent HST CANDELS study by Yano et al. (2016).  At
intermediate redshifts ($z\sim 1$) previous studies have found
conflicting results on the relationship between stellar age and the
compactness of passive galaxies, with indications that progenitor bias
may be playing a role at this epoch (Keating et al. 2015; Williams et
al. 2017).

Our findings may be explained if high-redshift post-starburst galaxies
are formed from the dramatic collapse of gas at high redshift, formed
from either a gas-rich merger (e.g.  Hopkins et al.  2009; Wellons et
al. 2015), or from gas inflow feeding a massive disc, which becomes
unstable and collapses by ``compaction'' (e.g. Dekel et al. 2009;
Zolotov et al. 2015; Tacchella et al. 2016). Star formation must then
be rapidly quenched, either by a central AGN or feedback from highly
nucleated star formation (e.g.  Hopkins et al. 2005; Diamond-Stanic et
al. 2012), leaving an ultra-compact post-starburst remnant. It may be
possible to test these evolutionary scenarios by comparing the
properties of post-starburst galaxies with their likely active
progenitors, i.e.  star-forming galaxies caught during the merging or
``compaction'' phase.  Current candidates include submillimetre
galaxies, many of which appear to be highly compact at
$\sim$250~$\mu$m in the rest-frame (e.g. Simpson et al. 2015), and the
high-redshift ``blue nuggets'' (e.g.  Barro et al.  2013; Mei et al.
2015; Barro et al. 2017).  Whatever the true progenitors for our PSBs,
the most likely explanation is that structural transformation occurred
immediately prior to quenching.  The fact that the PSBs and passive
galaxies have indistinguishable S\'{e}rsic indices (Figure
\ref{fig:sersicfig}) would suggest that most of the structural change
is already established when the star formation is quenched, unless the
structural transformation occurs on a much shorter timescale than the
$\sim$500 Myr lifetime for the PSB phase.

Following the formation of the proto-spheroid, there are currently two
leading explanations for the observed growth in passive galaxies with
cosmic time.  Minor gas-free mergers provide a plausible physical
mechanism (e.g. Bezanson et al. 2009; Naab, Johansson \& Ostriker
2009), and indeed there is evidence that high-redshift passive
galaxies are larger in dense environments, where such interactions are
more likely (e.g. Lani et al. 2013).  Alternatively, progenitor bias
may mimic the observed growth, since passive galaxies formed at later
times are typically larger (e.g. Poggianti et al. 2013; Carollo et
al. 2013).

Our finding that post-starburst galaxies are more compact than older
passive  galaxies, on average, would suggest that we are observing an
earlier phase in the lifetime of steadily-growing spheroids.  Thus
progenitor bias is unlikely to be the primary cause for the observed
growth at early times; our observed trends show
precisely the opposite behaviour (with younger galaxies being more
compact).  On the other hand, if massive post-starburst galaxies
represent newly-formed ``red nuggets'', it is notable that their
abundance is a strong function of redshift; massive (M$_{\ast}>
10^{10.5} ~$M$_{\sun}$) post-starburst galaxies are several times more
abundant at $z\sim 2$ compared to $z\sim 0.5$ (Wild et
al. 2016). Thus, while the majority of high-mass passive galaxies at
$z\sim 2$ are likely to have been through the ultra-compact
post-starburst phase (Wild et al. 2016), this may become an
increasingly less dominant channel towards lower redshifts, when
progenitor bias may play a more significant role in explaining size
evolution (e.g. see Fagioli et al. 2016, Williams et al. 2017).
Even at low redshift, however, there is evidence that the most compact
 quiescent galaxies  have evolved from
post-starburst progenitors (Zahid et al. 2016).

Finally, we note that the mass function for PSBs shows a very
distinctive evolution in shape (Wild et al. 2016).  At high redshift ($z\sim
2$) the mass function resembles that of quiescent galaxies, dominated
by high-mass systems with a sharp decline in space density above a
mass of M$_{\ast}\sim 10^{10.5} ~$M$_{\sun}$. At low
redshift ($z<1$) the population is dominated by lower-mass systems,
with a shape resembling the mass function for star-forming
galaxies. These features are interpreted as evidence for two distinct
formation channels for post-starburst galaxies; high-mass systems
formed from gas-rich dissipative collapse, and low-mass systems formed
from environmental quenching or the merging of normal disc galaxies (Wild et
al. 2016). Our structural findings are in good agreement with this
scenario, with the PSBs above the same characteristic mass displaying
distinctive, ultra-compact morphologies, consistent with a
highly-dissipative, gas-rich origin.  We will present a detailed study
of the structural parameters for low-mass PSBs in future work (Maltby
et al., in preparation).

\section{Conclusions}

\smallskip

We present a study of the structural parameters for a large sample of
photometrically-selected post-starburst galaxies in the redshift range
$1<z<2$, recently identified in the UKIDSS UDS field. These rare
transition objects provide the ideal sample for understanding the
links between the quenching of star formation and the structural
transformation of massive galaxies.

We demonstrate that deep ground-based near-infrared imaging can be
used to obtain robust sizes and S\'{e}rsic indices for large samples
of high-redshift galaxies.  From the resulting size--mass relation, we
find that massive (M$_{\ast}> 10^{10.5} ~$M$_{\sun}$) post-starburst
galaxies are exceptionally compact at $z>1$, with evidence that they
are more compact on average than established passive galaxies at the
same epoch.  Since most high-mass passive galaxies at $z>1$ are likely
to have been through a post-starburst phase (Wild et al. 2016), the
implication is that quiescent galaxies are most compact when they are
newly quenched, thereafter growing with cosmic time.  An important
caveat, however, is to acknowledge the considerable uncertainty in
stellar mass estimation, as discussed in Section
\ref{sec:uncertainties}. As an avenue for future research, it will be
important to determine whether stellar masses are systematically
overestimated for recently quenched stellar populations.

We also find that post-starburst galaxies show high S\'{e}rsic
indices, significantly higher than star-forming galaxies on average,
but statistically indistinguishable from the S\'{e}rsic indices of
established passive galaxies at the same epoch. We conclude that
massive post-starburst galaxies represent newly-formed compact
proto-spheroids.  Furthermore, the structural transformation of these
galaxies must have occurred before (or during) the event that quenched
their star formation.


\section*{Acknowledgements}

We thank Louis Abramson, Steven Bamford, Dale Kocevski, Mike Merrifield, and Ian Smail
for useful discussions.  We extend our gratitude to the staff at UKIRT
for their tireless efforts in ensuring the success of the UDS project.
We also wish to recognize and acknowledge the very significant
cultural role and reverence that the summit of Mauna Kea has within
the indigenous Hawaiian community.  We were most fortunate to have the
opportunity to conduct observations from this mountain.  V.W. and
K.R. and acknowledge support from the European Research Council
Starting Grant (PI Wild).  RJM acknowledges the support of the
European Research Council via the award of a consolidator grant (PI
McLure).









\appendix
\section{Robustness tests}
\label{sec:appendix}

\begin{figure*}
	\includegraphics[width=0.45\linewidth]{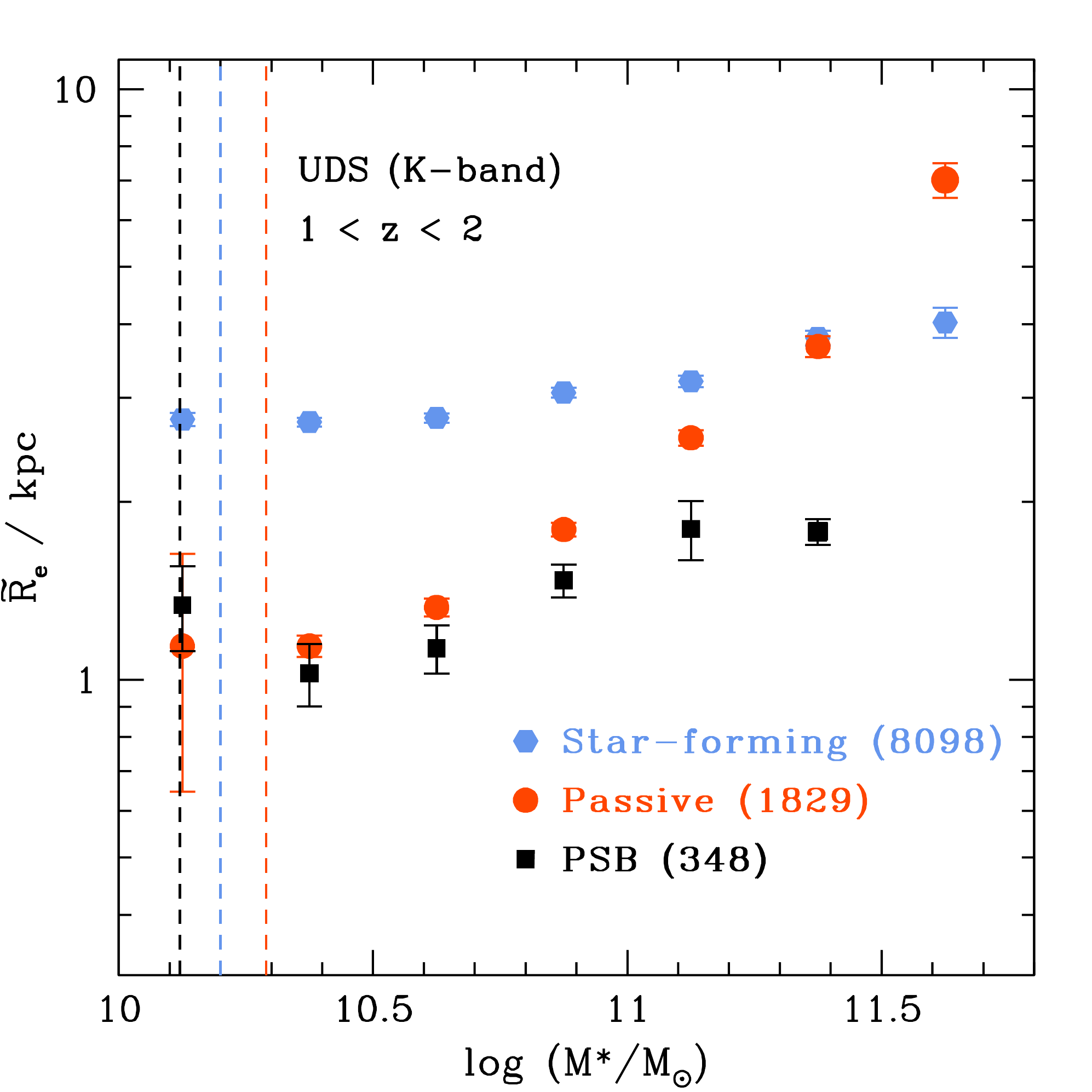}
	\includegraphics[width=0.45\linewidth]{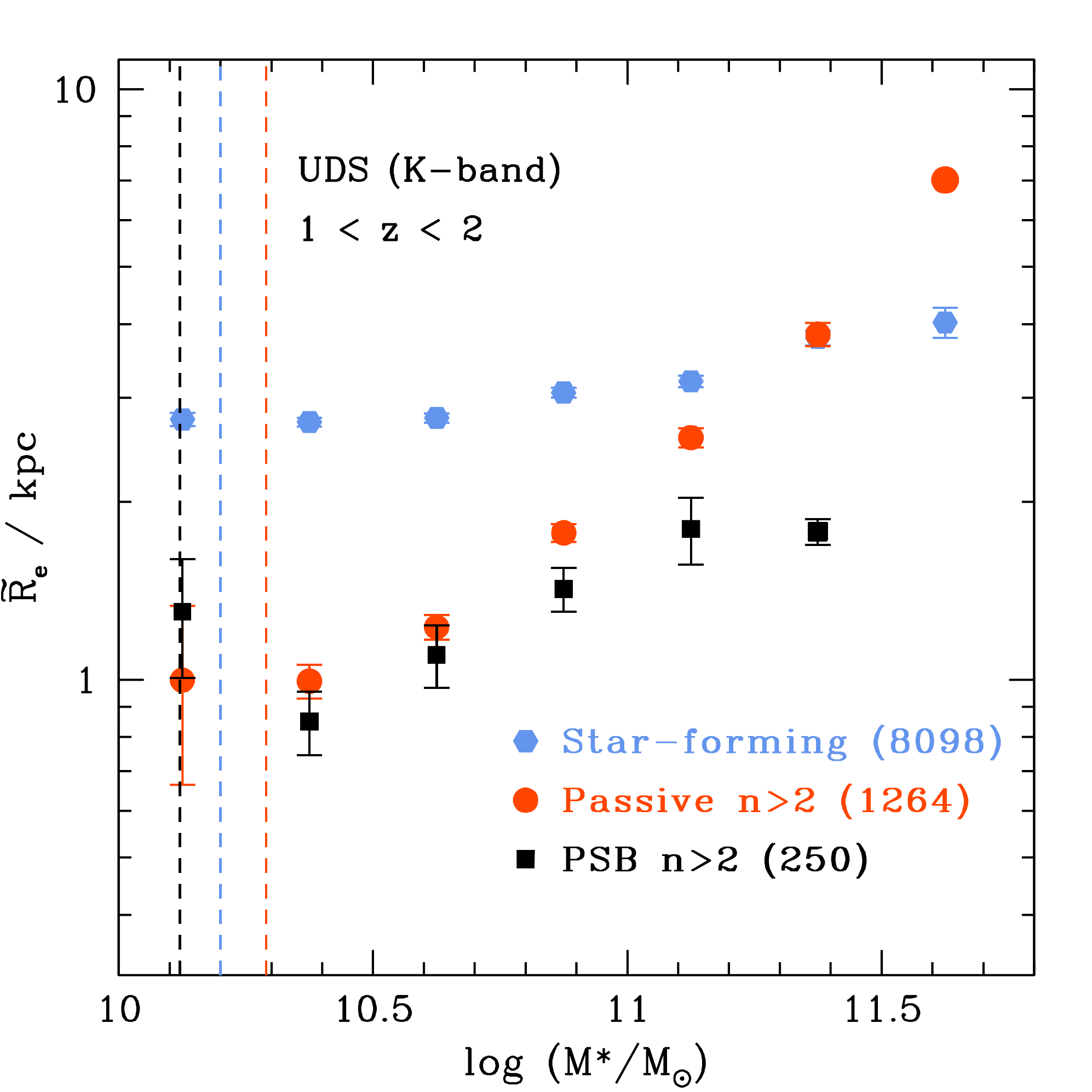}
    \caption{Two examples of tests for robustness.  The left panel
      shows the size--mass relation for comparison with  Figure \ref{fig:size-mass}
      ($1<z<2$), but using the median rather than the mean effective
      radius per bin.  The right panel shows the size--mass relation
      (also using median values) after the removal of all passive and
      post-starburst galaxies with $n<2$. Since star-forming galaxies
      have significantly lower S\'{e}rsic indices (see Figure
      \ref{fig:sersicfig}), this is a test to determine if
      contamination from SF galaxies is affecting the size--mass
      relations.  In both cases, we find that our primary conclusions
      are unchanged, i.e. post-starburst galaxies at high mass
      (M$_{\ast}> 10^{10.5} ~$M$_{\sun}$) appear significantly smaller
      than both star-forming and passive galaxies. 
}
    \label{fig:appenfig1}
\end{figure*}

In this section we discuss a number of additional tests that were
performed to investigate  the robustness of the size--mass relations presented
in this paper.

Throughout this work we have measured average sizes when comparing
galaxy populations, yielding evidence that post-starburst galaxies are
typically smaller than passive galaxies of comparable stellar
mass. The use of a mean may produce misleading results, however, if
either population is skewed by a significant number of outliers
(e.g. misclassified interlopers from other galaxy categories).  In
Figure \ref{fig:appenfig1} (left panel) we therefore reproduce the
size--mass relation from Figure \ref{fig:size-mass}, but this time
using median size values. In most bins the median sizes are slightly
smaller than the mean, as the size distributions show a slight tail to
high values. The significant differences between the passive and
post-starburst galaxies remain, however. At high-mass (M$_{\ast}>
10^{10.5} ~$M$_{\sun}$) the overall significance of the trends is
essentially unchanged compared to the results outlined in Section
\ref{sec:sizemass}. We conclude that the use of a mean has not biased
our primary conclusions.

As a further source of error, we considered the possibility that the
passive and post-starburst samples are contaminated by star-forming
galaxies, e.g. due to uncertainties in classification. If the
contaminating fraction is higher for the passive population (for
reasons unknown), this could skew the size measurements upwards.  We
tested for this effect by applying a cut in S\'{e}rsic index. Noting
the very different distributions in S\'{e}rsic index between
star-forming and passive samples (see Figure \ref{fig:sersicfig}), we
re-evaluated the size--mass relations using only passive and
post-starburst galaxies with $n>2$ (see Figure \ref{fig:appenfig1},
right panel), again using the median size to further reduce the impact
of interlopers.  The striking difference in sizes remains, with a
negligible reduction in significance.  Using a mean estimator yields
the same result, with a size--mass relation that is almost identical
to the upper-right panel of Figure \ref{fig:size-mass}.  We conclude
that ``contamination'' from galaxies with low S\'{e}rsic indices
(whether passive or star-forming) is not affecting our conclusions.

\begin{figure*}
	\includegraphics[width=0.45\linewidth]{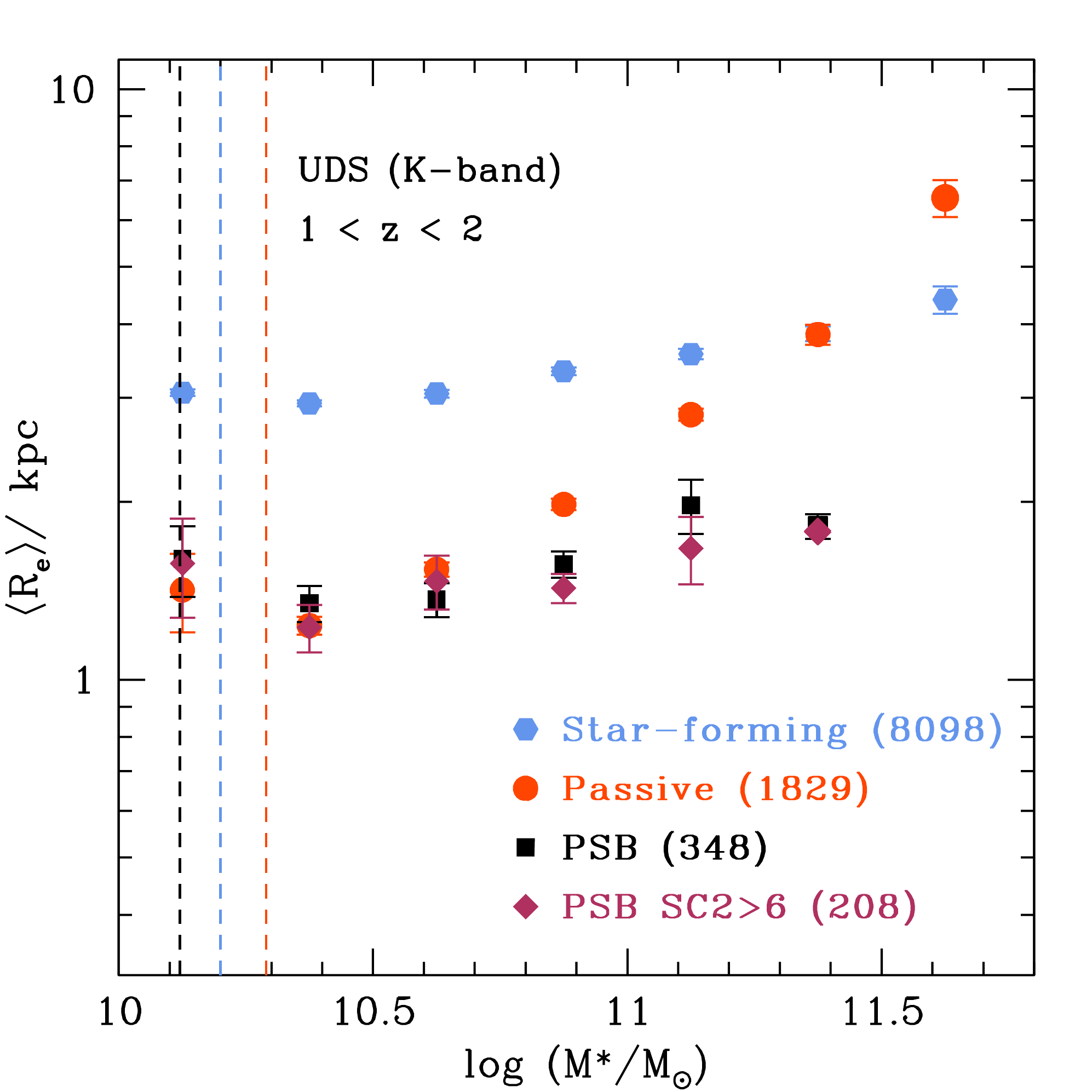}
	\includegraphics[width=0.45\linewidth]{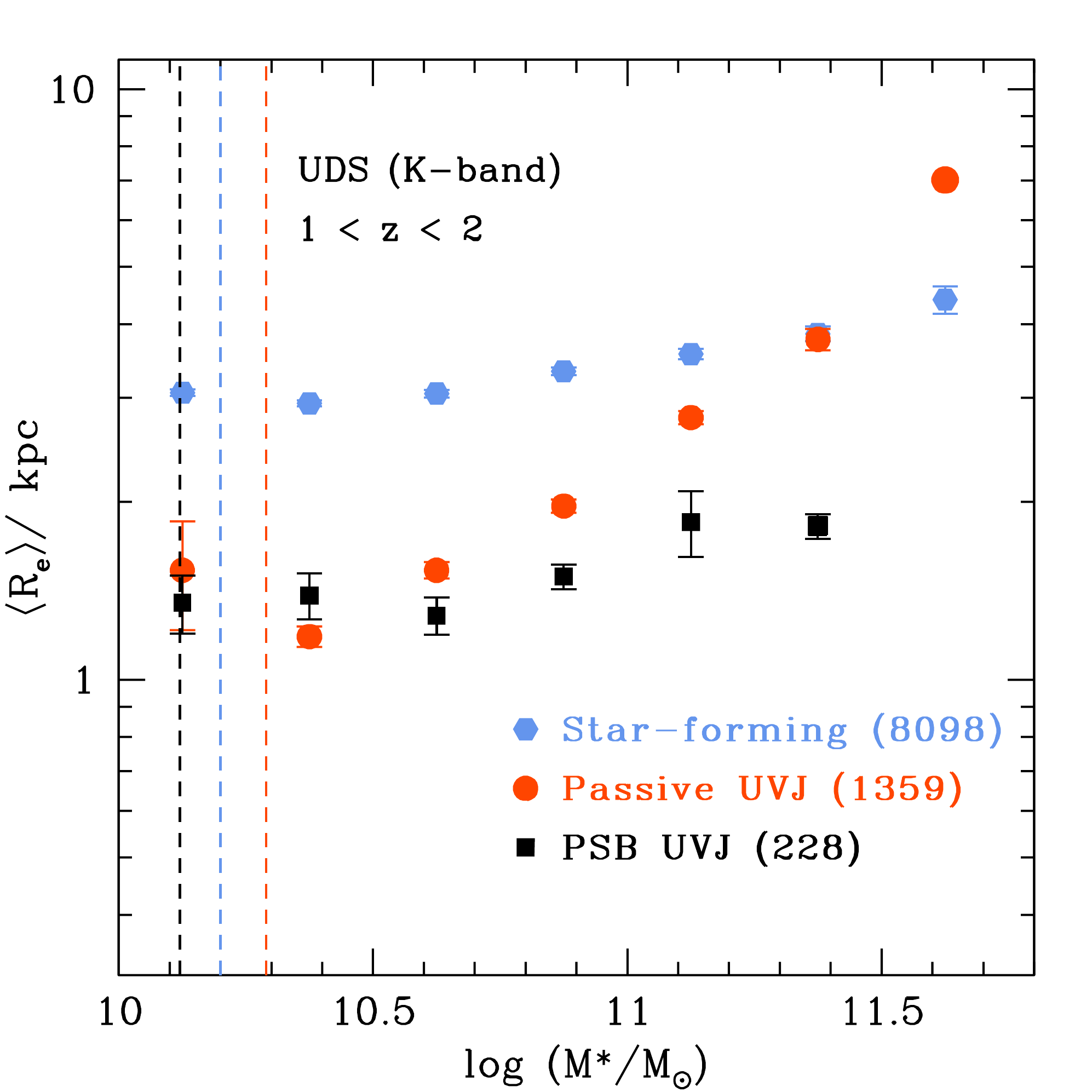}
    \caption{Two further tests for robustness. The left panel shows
      the ground-based size--mass relation, using the mean size (as
      shown in Figure \ref{fig:size-mass}), with the addition of a
      comparison sample of ``secure'' PSBs, identified further from
      the passive/PSB boundary (with supercolours $SC2$>6).  The right
      panel shows the size--mass relation using only the subset of
      passive galaxies and PSBs that simultaneously lie within the
      quiescent region of the UVJ diagram (e.g. see Figure
      \ref{fig:uvjfig}). In both cases, we find little change in the
      observed trends, with evidence for a slight enhancement
in the difference between high mass PSBs and passive
      galaxies using the restricted samples.}
    \label{fig:appenfig2}
\end{figure*}

As an additional test, we investigated the influence of using a
cleaner (though less complete) sample of post-starburst galaxies.  As
outlined in Section \ref{sec:intro} and Maltby et al. (2016), the
primary source of contamination is between post-starburst galaxies and
``normal'' passive galaxies. Depending on the precise selection
criteria, between 20--40\% galaxies in the PSB category would be
classified as passive (rather than PSB) using spectroscopy, while
6--10\% of the passive category would be spectroscopically classified
as PSBs (Maltby; private communication).  Based on Figure 3 in Maltby
et al. (2016), we therefore identify a ``cleaner'' PSB sample by
selecting galaxies further from the passive/PSB boundary, with
supercolours $SC2$>6. In this regime, formally 100\% of PSB candidates
(15/15) were confirmed with $W_{\rm H\delta}>5$\AA. Using this new
sub-sample, the resulting size--mass relation (evaluated using the
standard mean estimator) is shown in Figure \ref{fig:appenfig2} (left
panel).  We find that the difference in size compared to passive
galaxies remains, and in fact is slightly enhanced; the mean PSB
size is formally smaller in five out of six bins. We conclude that
contamination of the PSB category by older passive galaxies can only
act to  dilute the differences we observe.

As another test for contamination, we combined the PCA classification
with the classic UVJ criteria (e.g.  see Figure \ref{fig:uvjfig}), to
exclude PSBs and passive galaxies that are classified as
``star-forming'' using rest-frame UVJ colours.  The aim is to remove
red star-forming galaxies that may have been misclassified by the
PCA technique.  Using these joint criteria removes  26\%
of passive galaxies and 34\% of PSBs
from our primary sample ($z>1$,  M$_{\ast}>10^{10} ~$M$_{\sun}$).
 The resulting size--mass relations are shown in Figure \ref{fig:appenfig2}
(right panel). We find that the difference between the PSBs and
passive galaxies at high mass remains, and in fact is slightly
enhanced. 

Finally, we performed a variety of tests using more stringent cuts on
the structural parameters derived for our ground-based galaxy sample,
using the CANDELS dataset as a calibrating sample. No significant
differences were found. We noted, however,  that ground-based size
measurements become increasingly unreliable when GALFIT assigns a very
low axis ratio, $q<0.1$. Approximately 6\% of the galaxies in our
sample are affected, mostly among the star-forming galaxies, but also
affecting 3--4\% of the passive and post-starburst sample (mostly at
low mass; M$_{\ast}< 10^{10.5} ~$M$_{\sun}$). Removing these galaxies
had no major influence on the size-mass relations; in fact, the
difference in size between passive and post-starburst galaxies became
marginally more significant.

\bsp	
\label{lastpage}
\end{document}